\documentclass[aps,floatfix,twocolumn,a4paper,showpacs, nofootinbib,
superscriptaddress,10pt]{revtex4}
\usepackage{graphicx,float}\usepackage{graphicx,float}
\usepackage[all]{xy}
\usepackage{amsmath,upgreek}
\usepackage{amssymb}
\usepackage{color}
\usepackage{epsfig}		
\usepackage{graphicx,epstopdf}
\usepackage{subfigure}
\usepackage{pdfpages}
\usepackage[colorlinks,hyperindex]{hyperref}

\definecolor{green1}{RGB}{0,128,0} 
\hypersetup{hidelinks,backref=true,pagebackref=true,hyperindex=true,colorlinks=true,breaklinks=true,urlcolor= blue}
\hypersetup{%
  colorlinks = true,
  linkcolor  = blue,
  citecolor = green1,
}
\usepackage{bookmark,textgreek}
\usepackage{hyperref,color,xcolor}
\hypersetup{hidelinks,hyperindex=true,colorlinks=true,breaklinks=true,urlcolor= blue}
\hypersetup{%
  colorlinks = true,
  linkcolor  = blue
}

\newcommand{\bes}{\begin{subequations}}
\newcommand{\ees}{\end{subequations}}
\def\ben{\begin{eqnarray}}
\def\een{\end{eqnarray}}
\def\be{\begin{equation}}
\def\ee{\end{equation}}

\begin{document}

\title{Informational entropic Regge trajectories of meson families in AdS/QCD}
\author{A. E. Bernardini} 
\affiliation{Departamento de F\'isica, Universidade Federal de S\~ao Carlos,
PO Box 676, 13565-905, S\~ao Carlos, SP, Brazil}
\email{alexeb@ufscar.br}
\author{R. da Rocha}
\email{roldao.rocha@ufabc.edu.br}
\affiliation{Federal University of ABC, Center of Mathematics, Computing and Cognition, Santo Andr\'e, Brazil}\email{roldao.rocha@ufabc.edu.br}

\begin{abstract}
Bulk mesons propagating in chiral and gluon condensates, in a gravity background,
 are scrutinized in holographic soft wall AdS/QCD models, involving  deformed dilatonic backgrounds. The configurational entropy of the $a_1$ axial vector, the $\rho$ vector, and the $f_0$ scalar meson families is then computed.
Two types of {\color{black}{informational entropic}} Regge trajectories are then obtained, 
where the logarithm of the mesons configurational entropy is expressed in  terms of both the experimental meson mass spectra and their excitation number as well. Therefore the mass spectra of the next generation of elements in each meson family, besides being predicted as eigenvalues of Schr\"odinger-like equations, are estimated with better accuracy and discussed.
 \end{abstract}
\maketitle

\section{Introduction}
The Shannon's information entropy paradigm resides in encoding information 
in stochastic processes \cite{shannon}.
The configurational entropy (CE) is a quantity that implements the information entropy  as a measure that logarithmically evaluates the number of bits needed to designate the organization of a system. In particular, the CE comprehends the information compression into the configuration of wave modes  in a physical system \cite{Gleiser:2011di,Gleiser:2012tu}. 
Among the so called configurational information-measures \cite{Gleiser:2018kbq}, the CE encompasses the informational quantification of the spatial complexity of a localized system  \cite{Gleiser:2014ipa,Sowinski:2015cfa}. 
 The wave modes and particle excitations, that are correlated to 
 critical points of the CE, have been shown to be more dominant or abundant among all other modes and, hence, more detectable or observable in Nature  \cite{Bernardini:2016hvx,Bernardini:2016qit,Braga:2017fsb}.   The CE plays a relevant role 
 in the study of phase transitions, that are also driven by critical points of the CE, underlying diverse physical systems \cite{Gleiser:2014ipa,Sowinski:2017hdw,Sowinski:2015cfa,Braga:2016wzx}. A meticulous overview on the information entropy formalism can be 
  seen in Ref. \cite{Witten:2018zva}. 

Quantum chromodynamics (QCD) governs the strong interactions among gluons and quarks. 
The AdS/QCD holographic setup presents an AdS$_5$  bulk\footnote{Anti-de Sitter.}, wherein gravity, that is weakly coupled, emulates the dual setup to the 4d (conformal)  field theory (CFT), that is strongly coupled on the AdS$_5$ boundary. In the duality dictionary, physical fields in the AdS$_5$ bulk are dual objects to 4d  operators of QCD \cite{Natsuume:2014sfa}. It is worth to mention that the bulk fifth dimension is nothing more than the energy scale of the theory. Confinement can be then implemented either by a Heaviside cut-off in the bulk -- the hard wall \cite{Polchinski:2001tt,BoschiFilho:2002ta} --   or by a dilatonic field, 
that accomplishes a smooth cut-off along the AdS$_5$ bulk -- the soft wall model \cite{Csaki,Karch:2006pv}. From a phenomenological point of view,  the quark-gluon plasma (QGP), the mesonic mass spectra and their Regge trajectories  \cite{Brodsky:2014yha,Li:2013oda} and other quantities were derived, using the holographic soft wall AdS/QCD, being precisely corroborated by experimental data  \cite{pdg1}. Soft wall AdS/QCD models implement the (chiral) symmetry breaking \cite{Gkk,rold} and confinement as well  \cite{Karch:2006pv,zhang,sui1,Colangelo:2011sr}.   In particular, mesonic phenomenology can be thus allocated into the soft wall AdS/QCD. As shall be used 
throughout this paper, although QCD is driven by a SU(3) gauge symmetry, one can supersede it by a SU($N_c$) group. Hence, the so called 't Hooft large-$N_c$ limit \cite{hooft} plays an important role 
on the soft wall.

In the above discussed context, the CE has been recently promoted to a relevant setup in 
the context of holographic AdS/QCD models. Besides probing important informational aspects of the AdS/QCD, the CE also supports some foundations to better understand mesonic states in QCD phenomenology. 
Some of the existing experimental data regarding meson families and glueball states were corroborated by the CE, that points into the direction of the most abundant and dominant physical states in QCD, in an intense research program. In fact, holographic AdS/QCD  models were first scrutinized, in the CE framework, in Refs. \cite{Bernardini:2016hvx,Bernardini:2016qit,Braga:2017fsb}. \textcolor{black}{This subject was recently
extended to baryons and exotic states in Ref. \cite{Colangelo:2018mrt}}. Mesonic excitations with lower $s$-wave resonances were proved, in Ref. \cite{Bernardini:2016hvx}, to present dominance over their higher $s$-wave counterparts. Thereafter, the CE 
paradigm was employed to scrutinize scalar  glueball states in Ref. \cite{Bernardini:2016qit}, corroborating to lattice and experimental data \cite{pdg1}. In addition, the CE of dynamical AdS/QCD with tachyonic potentials improved our understanding about  mesons  phenomenology in Ref. \cite{Barbosa-Cendejas:2018mng}.  Moreover, the CE that underlies bottomonium and charmonium states at zero temperature supports the experimental rareness of quarkonia states with higher masses, as discussed in Ref.  \cite{Braga:2017fsb}. The finite temperature case was then implemented in Ref. \cite{Braga:2018fyc}, with unexpected and relevant quarkonia new features.  Finally, QGPs with topological defects were also explored in the CE setup  \cite{daSilva:2017jay}. 
The CE also improved the understanding of the AdS/QCD light-front wave function, where the 
color-glass condensate regime was used to study mesons in Refs. \cite{Karapetyan:2018oye,Karapetyan:2018yhm,Karapetyan:2016fai,Karapetyan:2017edu}. With the CE tools, heavy ion collisions were also studied   \cite{Ma:2018wtw}. 
Furthermore, aspects of the CE unraveled prominent features of the gravity side of AdS/QCD and related phenomena. The Hawking--Page phase transition was studied in Ref.  \cite{Braga:2016wzx} in the context of the CE. Posterior to the influential works    
 \cite{Gleiser:2013mga,Gleiser:2015rwa}, graviton condensates were explored with by the CE, in AdS/CFT membrane paradigm \cite{Casadio:2016aum}. The CE apparatus was further employed in field theory, in various contexts  \cite{roldao,Correa:2016pgr,Alves:2017ljt,Alves:2014ksa}.  
Here one big step further is aimed, besides corroborating to experimental data 
of $a_1$ axial vector, the $\rho$ vector, and the $f_0$ scalar mesons   families. 
Using both the quadratic and deformed dilatonic backgrounds, 
informational entropic Regge trajectories shall be derived, relating the logarithm of the CE to the $n$ excitation number of mesonic states, for each one of the meson families. 
The mesons mass spectra are well known 
to be predicted in the soft wall AdS/QCD, with good accuracy. In fact, for both the dilatonic backgrounds, the equations of motion (EOMs) of a graviton-dilaton-gluon action are equivalent to Schr\"odinger-like equations, whose eigenvalues consist 
of the meson squared mass spectra. Their eigenfunctions represent the mesonic states and excitations, for the $a_1$ axial vector, the $\rho$ vector, and the $f_0$ scalar mesons families. The CE for these meson families shall be also computed with respect to the meson families mass spectra, revealing a second type of  informational entropic Regge trajectories.  Hence, one can extrapolate the meson mass spectra 
from these informational entropic Regge trajectories, also predicting the mass of the next generation of elements in each meson family, corresponding to higher $n$ excitation numbers, with good accuracy.  The masses of the first mesonic excitations of the next generation, in each meson family, shall be then estimated and discussed. 
This article is devised as follows: Sect. \ref{ii} is devoted to briefly review the soft wall AdS/QCD framework, mainly emphasizing the standard quadratic dilaton model. The  mass spectra of 
$\rho$ vector, $a_1$ axial vector, and  $f_0$
scalar mesonic states shall be revised. Sect. \ref{iii} is dedicated to 
introduce the chiral and gluon condensates, with a two flavor system in the graviton-dilaton-gluon setup, for both the quadratic and deformed
dilatonic fields.
In Sect. \ref{ivi}, the CE is computed for the $a_1$ axial vector, the $\rho$ vector, and the $f_0$ scalar mesons families, as a function of the $n$ excitation number. Hence, informational entropic Regge trajectories are read off these calculations, 
showing a relation between the logarithm of the CE, for all regarded mesons families, and their $n$ excitation modes. Besides, the meson mass spectra for higher excitation numbers can be also extracted from  a second kind of informational entropic Regge trajectories, that relate the logarithm of the CE and the 
meson mass spectra, for each meson family. Hence, the mass spectra 
of higher $n$ excited mesonic states, in each meson family, are estimated with good accuracy.  In Sect. \ref{iv}, our concluding remarks, outlook, and perspectives are drawn
\vspace*{-.5cm}
\section{soft wall AdS/QCD}
\label{ii}
The AdS$_5$ vacuum bulk has a 4d boundary, that supports a gauge theory, that is conformally invariant,  emulating the standard QCD when $N_c\gg 1$. 
The boundary conformal symmetry can be broken, making QCD to describe the confinement. In this regime, gravity in the bulk is dual to the QCD at the boundary. A straightforward way to break the boundary symmetry is, for example, to endow the  AdS bulk with a dilatonic  field. 
Since QCD approximately recovers conformal symmetries, in a high energy regime, 
then the pure  AdS bulk must prevail in the ultraviolet (UV) regime.  
Mesons families can be emulated in the holographic soft wall  AdS/QCD   \cite{Karch:2006pv,Braga:2015jca} and in its extended versions  
\cite{Colangelo:2008us,Gkk,sui1,Afonin:2012jn}, including the dynamical  models \cite{Rougemont:2017tlu,dePaula,dePaula:2009za,Capossoli:2016ydo,Barbosa-Cendejas:2018mng} and scalars and vector mesonic states \cite{Ihl:2010zg}. 
The Regge trajectories for excited light-flavor mesons were originally derived in 
Ref. \cite{Karch:2006pv}, on a soft wall model endowed with a 
quadratic dilaton, $\Upphi(z)=\mu^2z^2$,  where  $\mu$ introduces an energy scale in QCD \cite{Batell:2008me}. 
The ${\rm AdS}_5$ background  bulk metric is expressed, in conformal coordinates, as  
\begin{equation}\label{bulkm}
ds^{2}=g_{mn}dx^{m}dx^{n}=e^{2{\rm A}(z)}\!\left( \upeta_{\mu\nu}dx^{\mu}dx^{\nu}+dz^{2}\right),
\end{equation}
 for the warp factor ${\rm A}(z)=-\log(z/\ell)$, where the $ \upeta_{\mu\nu}$ denote the 4d space-time metric components and $\ell$ is related to the bulk curvature radius. Other warp factors, extending the soft wall AdS/QCD, were used in Refs. \cite{dePaula,dePaula:2009za,BallonBayona:2007qr}. 
Hereon $m,n,q$ denote bulk indexes, running from 0 to 4, where $x^m = (x^\mu, x^4),$ for $x^\mu$ denoting 4d space-time coordinates and $x^4$ denoting the bulk coordinate.  Light-flavor mesonic excitations  are represented by bulk $\mathfrak{X}(z)$ fields, that are dual objects to the quark-antiquark operator, with associated mass $m_\mathfrak{X}$, and 
 governed by the action \cite{Karch:2006pv}
\begin{eqnarray}
 S=-\int \, e^{-\Upphi(z)} \sqrt{-g}\,{\rm Tr}\,\mathfrak{L}\,d^5x, \label{softw}
 \end{eqnarray} 
where  
\begin{eqnarray}\label{lagran}
\!\!\!\!\!\mathfrak{L}=D_m\mathfrak{X}D^m\mathfrak{X}+m_\mathfrak{X}^2 \mathfrak{X}_m\mathfrak{X}^m
 +\frac{N_c}{48\pi^2}\left(F_R^2+F_L^2\right),
 \end{eqnarray} 
where the $A_L^{m}$ and
$A_R^{m}$ gauge fields drive the SU(2)$_L\times$ SU(2)$_R$ chiral flavor symmetry of QCD. Each SU(2), with $\{f_b/2\}$ ($b=1,2,3$) generators, correspond to one quark flavor. The left and right gauge field strengths respectively read 
\begin{eqnarray}
F_{L,R}^{mn}&=&\partial^{[m}{A_{L,R}^{n]}}-i[A_{L,R}^{m},A_{L,R}^{n}],\end{eqnarray}
where $A_{L,R}^{m}= A_{L,R}^{ma} f_a$. The covariant derivative is explicitly given by 
$D^m \mathfrak{X}=\partial^m \mathfrak{X}-i A_L^m \mathfrak{X}+i\mathfrak{X} A_R^m.
$ 
The $\mathfrak{X}(z)$  field  incorporates the $S$ (scalar) and $P$ (pseudo-scalar) fields,
as \cite{Li:2013oda}
\begin{equation}
\mathfrak{X}(z) = \left(S+{\upxi(z)}\right)\exp\left(iP^{b}t^{b}\right),
\label{scalarfield}
\end{equation}
where $\upxi(z)$ is a vacuum expectation
value that breaks chiral symmetry \cite{Li:2013oda}. 
To describe the vector and axial vector meson, the left, $A_L$, and right, $A_R$, gauge
fields can be split into the vector ($V$) and axial vector ($\mathring{A}$) fields, as  \cite{Li:2013oda} 
\begin{subequations}
\begin{eqnarray}
V^m&=&(A^m_R+A^m_L)/2\label{vv12}\\
\mathring{A}^m&=&(A^m_R-A^m_L)/2,\label{a12} 
\end{eqnarray}
\end{subequations}
yielding the respective gauge field strengths, 
\begin{eqnarray}
F_{V}^{mn}&=&\partial^{[m}{V^{n]}}-{i}[V^{m},V^{n}],\\
F_{\mathring{A}}^{mn}&=&\partial^{[m}{\mathring{A}^{n]}}-{i}[\mathring{A}^{m},\mathring{A}^{n}].
\end{eqnarray}
With respect to the vector $V$ and axial vector $\mathring{A}$ fields, the soft wall Lagrangian (\ref{lagran})
reads 
\begin{equation}
\!\!\!\!\!\!\!\mathfrak{L}={\rm D}_m\mathfrak{X}{\rm D}^m\mathfrak{X}+m_\mathfrak{X}^2 \mathfrak{X}_m\mathfrak{X}^m
 +\frac{N_c}{24\pi^2}\left(F_{\mathring{A}}^{2}+F_{V}^{2}\right),
\label{softwvecax}
\end{equation}
for ${\rm D}_m=\partial_m\mathfrak{X}+i(XV_m-V_mX-A_mX-XA_m)$.  
The EOM for the $\upxi(z)$ field then reads 
\begin{eqnarray}
\!\!\!\!\!\!\!\!\upxi''(z)\!+\!(3 {\rm A}'(z)\!-\!\Upphi'(z))\upxi'(z)\!-\! m_\mathfrak{X}^2(z) e^{2{\rm A}(z)}\upxi(z)\!=\!0.
\label{vevo}
\end{eqnarray}
In the standard soft wall model, $\Upphi(z)=\mu^2 z^2$ \cite{Karch:2006pv}, and Eq. (\ref{vevo}) has solutions 
\begin{eqnarray}
\!\!\!\!\!\!\upxi(z)\!=\!c_1 z^3\exp\left({z^2}/{2}\right)  I(z)\!+\!c_2{\rm G}_{12}^{20}\left(\!-z^2\Big\vert
\begin{array}{c}
 1 \\
 {1}/{2}, {3}/{2}
\end{array}
\!\right) ,
\end{eqnarray}
with $I(z)=I_0\left(\frac{z^2}{2}\right)
+I_1\left(\frac{z^2}{2}\right)$, where $I_n(z)$ denotes the first kind modified Bessel function, and the second term is  the Meijer function of $\{1,2;2,0\}$ order.
Let one denotes by ${\tt v}_n$ the functions that describe the $\rho$ vector mesons,
and by ${\tt a}_n$ those ones that represent the $a_1$ axial vector mesons, whereas 
the ${\tt s}_n$ denote the functions that comprise the $f_0$ scalar mesons. 
For the vector ($\rho$), axial vector ($a_1$), and scalar ($f_0$) meson families, for $n=1,2,\ldots$, the meson spectra in the standard soft wall AdS/QCD model are ruled by the following EOMs\footnote{\textcolor{black}{At this point, a clarifying note relative to the Eq.~(14) from Ref.~\cite{Karch:2006pv} is necessary. For mode functions ${\tt v}_n(z)$ identified by the quantum mechanical wave functions, $\psi_n(z)$, i.e. ${\tt v}_n(z)=\exp(z^2/2)\,z^{1/2}\psi_n(z)$, in the literature, one frequently notices that $n=1$ refers to the ground state of the quantum system (as to be a clear reference to the Hydrogen atom quantum mechanics, being a standard notation in the literature).
Therefore, the right expression for $\psi_n(z)$ (a corrected one from Ref.~\cite{Karch:2006pv}) should be given by}
\textcolor{black}{\begin{equation}
\psi_n(z) = e^{-z^2/2}\,z^{m+1/2}\sqrt{\frac{2(n-1)!}{(m+n-1)!}}L^m_{n-1}(z^2),
\end{equation}
such that $I_\psi\equiv\int_{0}^{\infty}\!\!dz\,\psi^*_n(z)\psi_n(z)$ reads
\begin{eqnarray}
\!\!\!\!\!\!\!\!\!\!I_\psi\!\!&\!\!=\!\!&\!\! \frac{2n!}{(m\!+\!n\!-\!1)!}\int_{0}^{\infty}\!\!\!dz\,e^{-z^2}z^{2m+1}L^m_{n-1}(z^2)\,L^m_{n-1}(z^2)\nonumber\\
\!&\!=\!&\!\! \frac{2n!}{2(m\!+\!n\!-\!1)!}\int_{0}^{\infty}\!\!\!\!dre^{-r}r^{m}L^m_{n-1}(r) L^m_{n-1}(r)= 1,
\end{eqnarray}
so as to have $n=1$ for a typical ground state (in correspondence with the Hydrogen-like radial quantum number).}} \begin{eqnarray}
\left(-\frac{d^2}{dz^2}+V_{\tt s}(z)\right){\tt s}_n(z)&=&m_n^2 {\tt s}_n(z),\label{eomf0} \\
\left(-\frac{d^2}{dz^2}+V_{\rho}(z)\right){\tt v}_n(z)&=&m_n^2 {\tt v}_n(z),\label{eomrho} \\
\left(-\frac{d^2}{dz^2}+V_{\tt a}(z)\right){\tt a}_n(z)&=& m_n^2 {\tt a}_n(z), \label{eoma1}
\end{eqnarray}
 where the Schr\"odinger-like potentials are respectively given by 
\begin{eqnarray}
\!\!\!\!\!\!\!\!\!\!\!\!V_{\tt s}(z)&=&\frac{(\Upphi'(z)-3{\rm A}'(z))^2-1}{4}-m_\mathfrak{X}^2e^{2{\rm A}(z)}.\label{v12}\\
\!\!\!\!\!\!\!\!\!\!\!\!V_{\rho}(z)&=&\frac{(\Upphi'(z)-{\rm A}'(z))^2-1}{4}, \label{v10}\\
\!\!\!\!\!\!\!\!\!\!\!\!V_{\tt a}(z)&=& \frac{(\Upphi'(z)-{\rm A}'(z))^2-1}{4}+\frac{48\pi^2}{N_c} e^{2{\rm A}(z)} \upxi^{2}(z),\label{v11}
\end{eqnarray}

\begin{figure}[H]
\centering
\includegraphics[width=8.2cm]{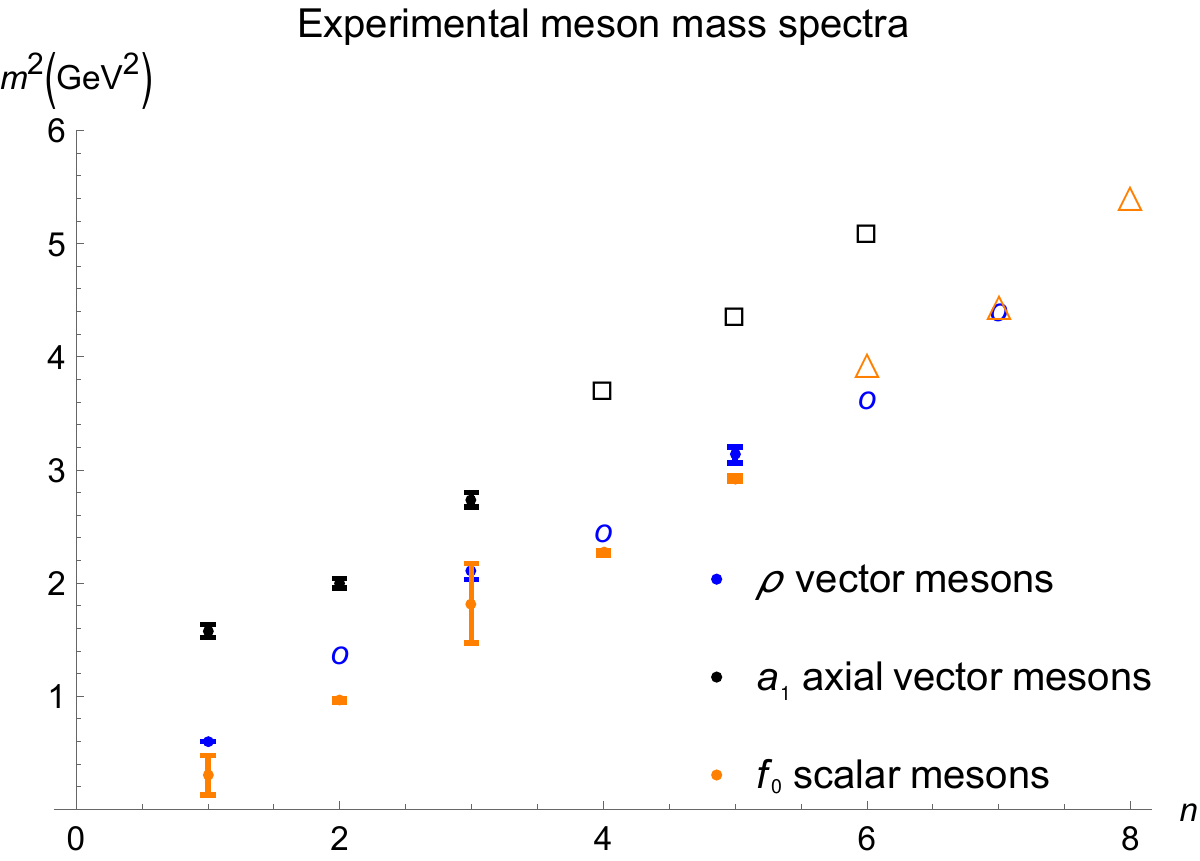}
\caption{\textcolor{black}{Experimental mesonic states mass spectra, as a function of the 
$n$ excitation number. The orange points represent the $f_0(500)$, $f_0(980)$, $f_0(1370)$, $f_0(1500)$, $f_0(1710)$ scalar mesons. The $f_0(2020)$, $f_0(2200)$ and $f_0(2330)$ states, omitted from the summary table in PDG, are simbolyzed by triangles (``${\tiny\Delta}$''). The blue points depict the $\rho(770)$, $\rho(1450)$, $\rho(1700)$  vector meson states, whereas
the $\rho'(1450)$, $\rho(1570)$, $\rho(1900)$, $\rho(2150)$ and $\rho(2270)$ are depicted by circles (``${\mathit\circ}$''). The black points represent $a_1(1260)$, $a_1(1420)$, $a_1(1640)$ axial vector mesons and the $a_1(1930)$, $a_1(2095)$ and $a_1(2270)$, omitted from the summary table in PDG, are depicted by empty boxes ``${\tiny \Box}$''.  All the experimentally established states have error bars \cite{pdg1}}.}
\label{f1}
\end{figure}

The potentials (\ref{v12}) -- (\ref{v11})  respectively determine the mass spectra of
the  $a_1$, $\rho$, and $f_0$ mesons families, as (squared) eigenvalues in Eqs. (\ref{eomf0})  -- (\ref{eoma1}). 
The experimental data for the $a_1$ axial vector, the $\rho$ vector, and the $f_0$ scalar mesons are shown in
Fig. \ref{f1}. 
\textcolor{black}{One should also notice that the $\rho$ states depicted in Fig. \ref{f1} regard the $\rho(770)$, $\rho(1450)$ and $\rho(1700)$ as confirmed
vector meson states in PDG, where between 500K and 1.98M events 
have been run. On the other hand, the $\rho(1570)$, $\rho(1900)$, $\rho(2150)$ and
$\rho(2270)$ are not established particles yet and
therefore are omitted from the summary table in PDG \cite{pdg1}. In fact,  the $\rho(1570)$ may be an  Okubo--Zweig--Iizuka-violating \cite{okubo}  decay mode  of the $\rho(1700)$ state. Together with the  $\rho(1900)$, $\rho(2150)$ and
$\rho(2270)$, they are listed in Ref. \cite{pdg1} as light unflavored mesons, with just 54 events already run. 
%\%
In addition,  the $f_0(500)$, $f_0(980)$, $f_0(1370)$, $f_0(1500)$, $f_0(1710)$ are established scalar meson particles  plotted in Fig. \ref{f1}, whereas the $f_0(2020)$, $f_0(2200)$ and $ f_0(2330)$ scalar mesonic states are still left out the summary table in PDG.  Finally, the $a_1(1260)$, $a_1(1420)$, $a_1(1640)$ axial mesons have been experimentally confirmed, whereas the  $a_1(1930)$, $a_1(2095)$ and $a_1(2270)$ axial vector mesonic states are also omitted from the summary table in PDG \cite{pdg1}, with just few events registered.}
%
%\vspace*{-.9cm}
\section{Two flavor soft wall AdS/QCD in graviton-dilaton-gluon bulk}
\label{iii}
\begin{figure}[H]
\centering
\includegraphics[width=6cm]{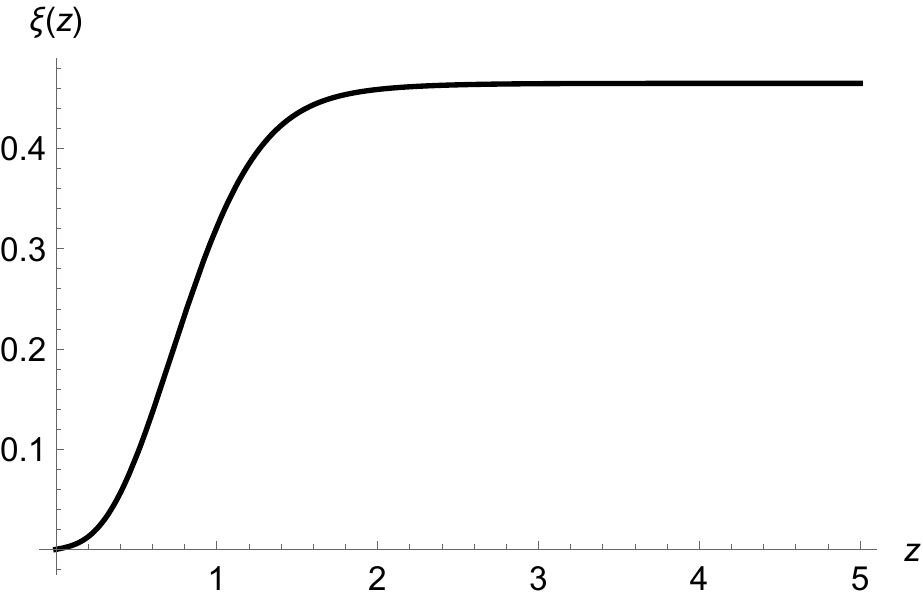}\medbreak
\caption{$\upxi(z)$ field, in the (\ref{quadraticd}) quadratic dilatonic background. }
\label{solucoes}
\end{figure}

\begin{figure}[H]
\centering
\includegraphics[width=6cm]{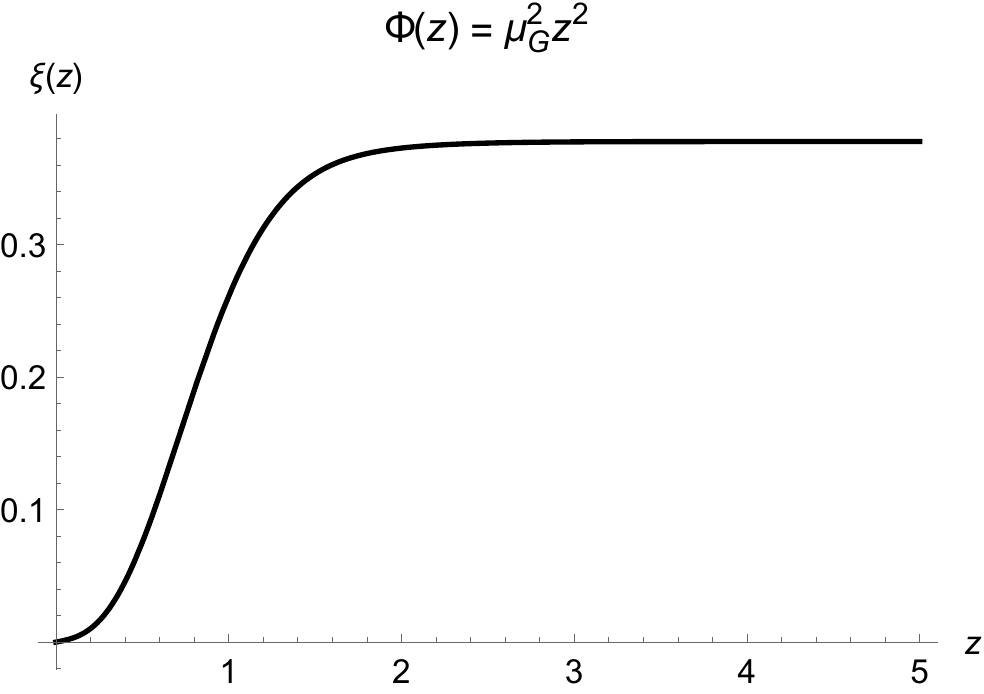}\\
\caption{$\upxi(z)$ field, in the (\ref{tanh}) deformed dilatonic background. }
\label{solucoes3}
\end{figure}

A two flavor soft wall AdS/QCD can be then considered, where a dilatonic background field is assumed in a chiral and gluon condensate background, with gravity \cite{Colangelo:2011sr}. For the pure gluon system, the scalar glueballs CE was already studied in Ref. \cite{Bernardini:2016qit}. Besides, Ref. \cite{Li:2013oda} used two types of dilaton background fields, yielding the glueball spectra in full compliance to lattice data, namely
\begin{eqnarray}
\Phi_1(z)&=&\mu_{\rm G}^2z^2, \label{quadraticd} \\
 \Phi_2(z)&=&\mu_{\rm G}^2z^2\tanh\left(\frac{\mu_{{\rm G}^2}^4z^2}{\mu_{\rm G}^2}\right).
\label{tanh}
\end{eqnarray}
The dual dimension-2 [dimension-4] gluon condensate has $\mu_{\rm G}$ [$\mu_{\rm G^2}$] energy scale.  The $\Phi_1(z)$ dilatonic field in Eq. \eqref{quadraticd} yields the meson spectra and also implements the quarks  confinement \cite{Karch:2006pv}. It is the dual object to 
the  gluon condensate with dimension-2, meaning the Bose--Einstein condensate consisting 
of strongly coupled paired gluons \cite{Gkk,Afonin:2012jn,Csaki}. 
The $\Phi_2(z)$ dilatonic field in Eq. (\ref{tanh}), at the UV regime behaves as $\lim_{z\rightarrow0}\Phi_2(z)=\mu_{{\rm G}^2}^4 z^4,$ being dual to a gluon condensate that has dimension-4 
\cite{xu}. At the IR regime,  $\lim_{z\rightarrow\infty}\Phi_2(z)=\mu_{\rm G}^2 z^2.$
A graviton-dilaton-gluon bulk action can be expressed as a sum of an Einstein-Hilbert action for pure gravity in the bulk, an action for gluons written with respect to the $\Upphi$ dilaton field, and an action for two flavor bulk mesons on a dilatonic background. This last part of the action implements 
the dynamics of the $\upxi(z)$ scalar field in Eq. (\ref{scalarfield}). The effective graviton-dilaton-gluon bulk action reads \cite{Li:2013oda}, 
\begin{eqnarray}\label{alll}
\!\!\!\!S\!=\!\kappa_5^2 \int \, \sqrt{-g}e^{-2\Upphi}\Big\{ \left[{R}/{4}+g_{mn}\partial^m\Upphi
\partial^n \Upphi - V_g(\Upphi) \right.\nonumber\\\left.
 \!\!\!\!\!\!\!\!\!\!\!\!\!\!\hspace*{-2.3cm}- 4\lambda e^{-\Upphi}\!\left(g_{mn}\partial^m\upxi \partial ^n \upxi
\!+\! V(\Upphi,\upxi)\right)\right]\!\Bigg\}\,d^5 x,
\end{eqnarray}
where $\lambda$ denotes a general coupling, and 
$V_g$ denotes the gluon system potential. The action (\ref{alll}) yields the following EOM: 
\begin{subequations}
\begin{eqnarray}\label{eomall1}
 \!\!\!\!\!\!\!\!\!\!\!\!\!\!\!-3{\rm A}''\!+\!3{\rm A}'^{2}\!+\!2\Upphi''\!-\!4{\rm A}'\Upphi'
 \!-\!2\lambda e^{\Upphi}\upxi'^{2}\!=\!0,&& \label{equacoes} \\
  \!\!\!\!8\Upphi''+24{\rm A}'\Upphi'-16\Upphi'^2\!-\!6\lambda\upxi'^{2}e^{\Upphi}\qquad\qquad\qquad
\nonumber&&\\  \!\!\!\!-3\frac{\partial}{\partial{\Upphi}}\left(\lambda e^{{7}\Upphi/3}V(\Upphi,\upxi)+V_g(\Upphi)
\right)e^{-4\Upphi/3+2{\rm A}}=0,&& \label{mall}\\
 -\upxi''\!+\!(\Upphi'\!-\!3{\rm A}')\upxi'\!+\!e^{2{\rm A}}\frac{\partial^2V(\Upphi,\upxi)}{\partial\upxi\partial\Upphi}\!=\!0.&& \label{eomall3}
\end{eqnarray}
\end{subequations}
In what follows $N_f$ stands for the number of flavors. 
In the UV regime, Refs. \cite{Li:2013oda,Cherman:2008eh} show that 
$
\lim_{z \rightarrow 0}\upxi(z)=\frac{m_q \uptau z}{2}+\frac{\upvarsigma}{2\uptau} z^3,$
 where $m_q$ is the quark mass,  $\upvarsigma$ denotes the string tension 
gluing the quark-antiquark condensate, and  $\uptau^2=\frac{N_c^2}{4\pi^2N_f}$, with $N_c=3$ and $N_f=2$. 
Ref. \cite{Li:2013oda} scrutinized   the heavy quark potential under the graviton-dilaton-gluon background (\ref{alll}), showing that the potential in (\ref{alll}) and (\ref{eomall3}) reads $V(\Upphi,\upxi)\approx \upxi^2\Upphi^2$, 
$\lim_{z \rightarrow \infty}{\rm A}'(z) = 0$ and $
\lim_{z \rightarrow \infty}{\rm A}(z) =a,$ 
for $a$ constant. 
%The equation of motion (\ref{equacoes}) in the IR  yields 
%$
%\lim_{z \rightarrow \infty}\upxi(z) =\mu_{\rm G} \sqrt{\frac{8}{\lambda}} e^{-\Upphi/2}.
%$ 
For both the $\Upphi_1(z)$ and $\Upphi_2(z)$, respectively in Eqs. (\ref{quadraticd}) and (\ref{tanh}), the parameters 
$\mu_{\rm G}= 0.43=\mu_{{\rm G}^2}$ were adopted in Ref. \cite{Li:2013oda}, in order to fit the  meson spectra data with good accuracy to experimental data. 
Besides, $\upvarsigma\approxeq 5.841\times 10^6\,{\rm MeV}$  and $m_q = 5.81$ MeV, for the $\Upphi_1(z)$ quadratic dilaton, whereas  $\upvarsigma\approxeq4.484\times 10^6\,{\rm MeV}$ and $m_q = 8.38$ MeV, for the $\Upphi_2(z)$ deformed dilaton. 
Numerical analysis of Eqs. (\ref{eomall1}) -- (\ref{eomall3}) in Ref. \cite{Li:2013oda} yield the solutions for $\upxi(z)$,  for  both the dilatonic backgrounds, in 
Figs. \ref{solucoes} and \ref{solucoes3}, in full compliance with the UV and the IR regimes for the dilaton fields in Eqs. (\ref{quadraticd}) and (\ref{tanh}).
\noindent 
\vspace*{-1.1cm}
\section{Informational entropic Regge trajectories and meson mass spectra}
\label{ivi}
\begin{figure}[H]
\centering
\includegraphics[width=7.5cm]{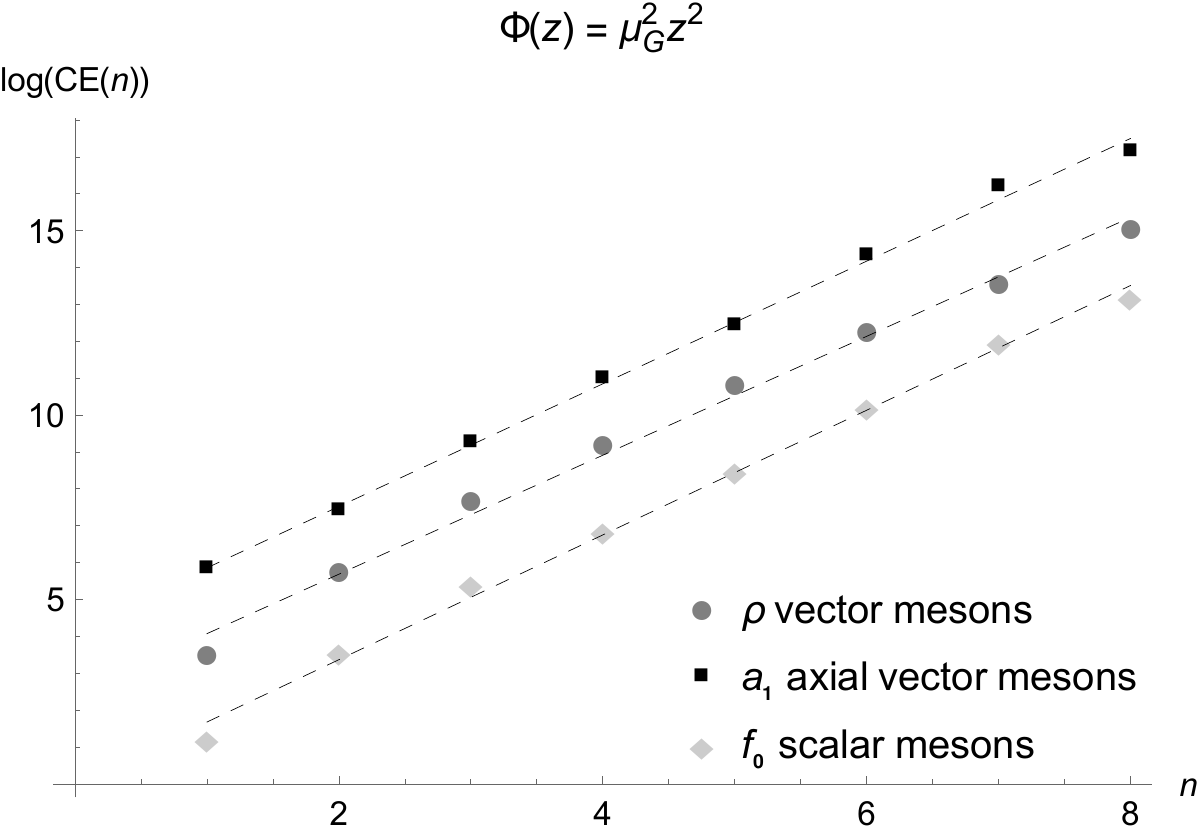}
\caption{Logarithm of the configurational entropy of mesons, the families of $\rho$, $a_1$, and $f_0$ mesons, in the   quadratic dilaton soft wall model.}
\label{f2}
\end{figure}

\begin{figure}[H]
\centering
\includegraphics[width=7.6cm]{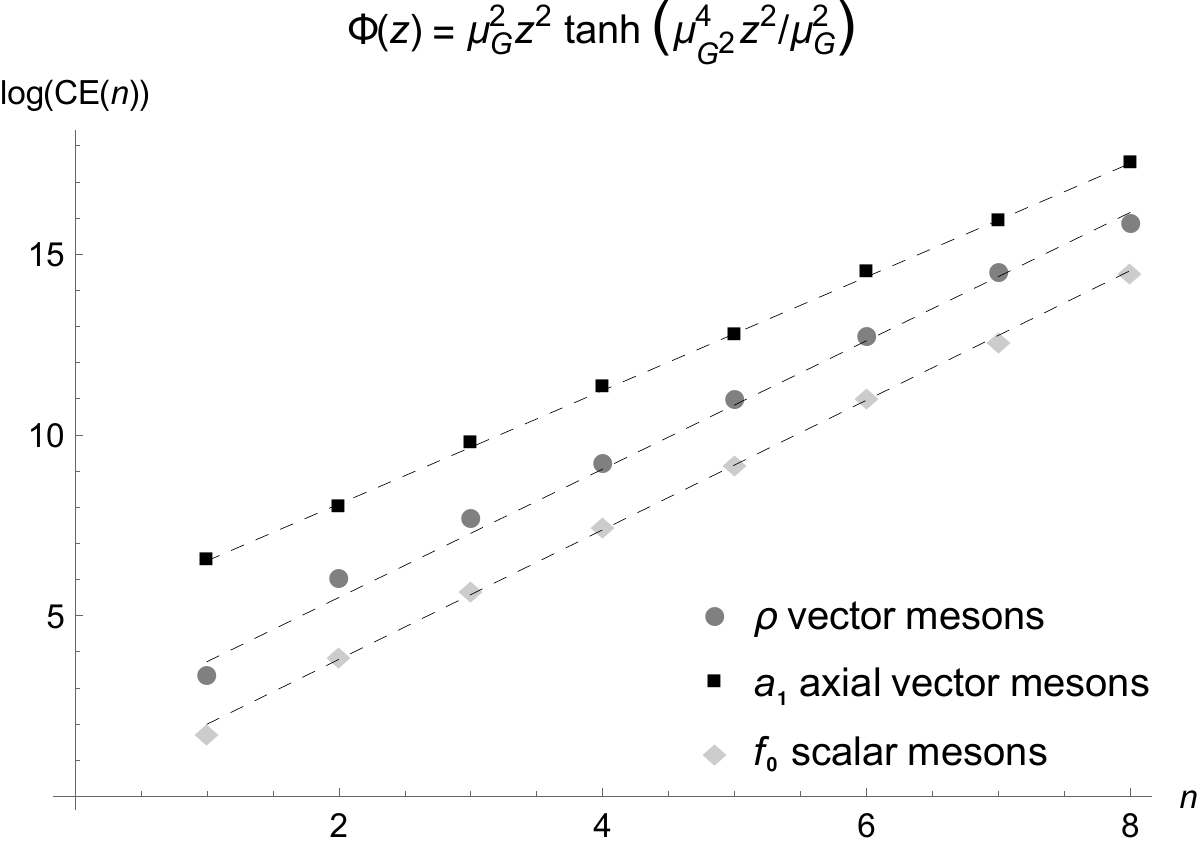}
\caption{Logarithm of the configurational entropy of the families of $\rho$, $a_1$, and $f_0$ mesons, in the deformed soft wall AdS/QCD.}
\label{f3}
\end{figure}

\begin{figure}[H]
\centering
\includegraphics[width=7.3cm]{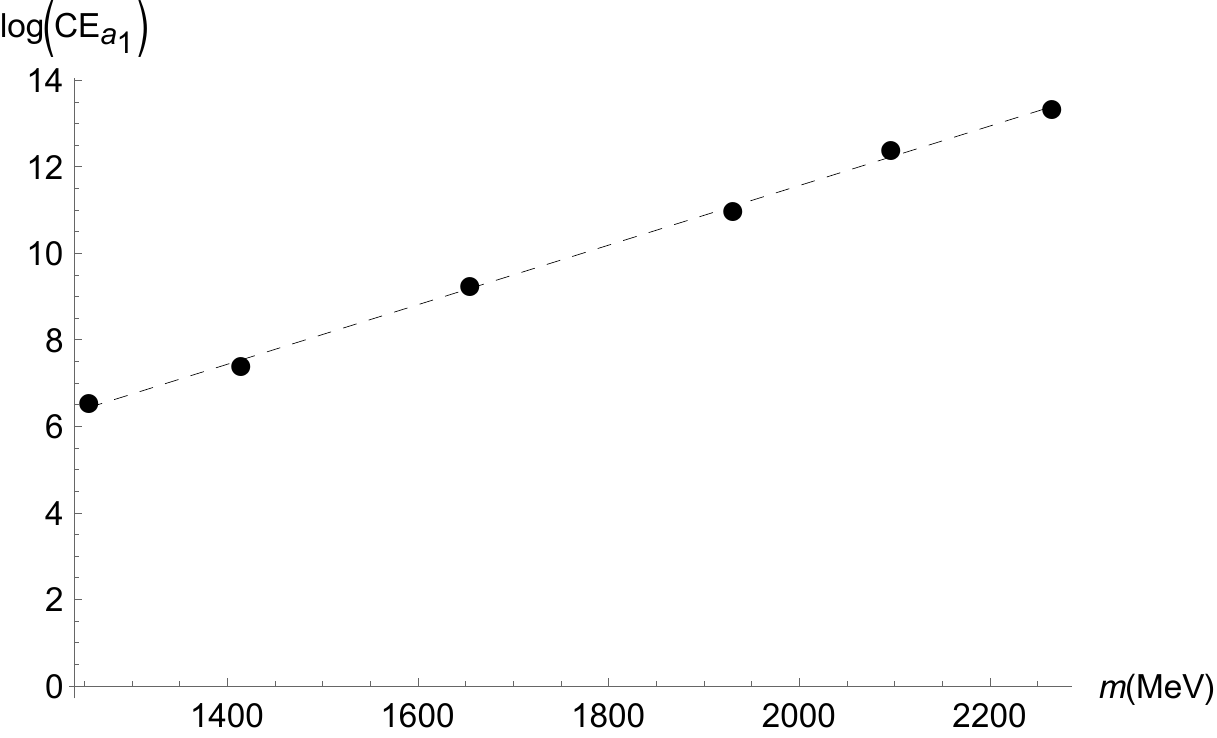}\medbreak
\includegraphics[width=7.1cm]{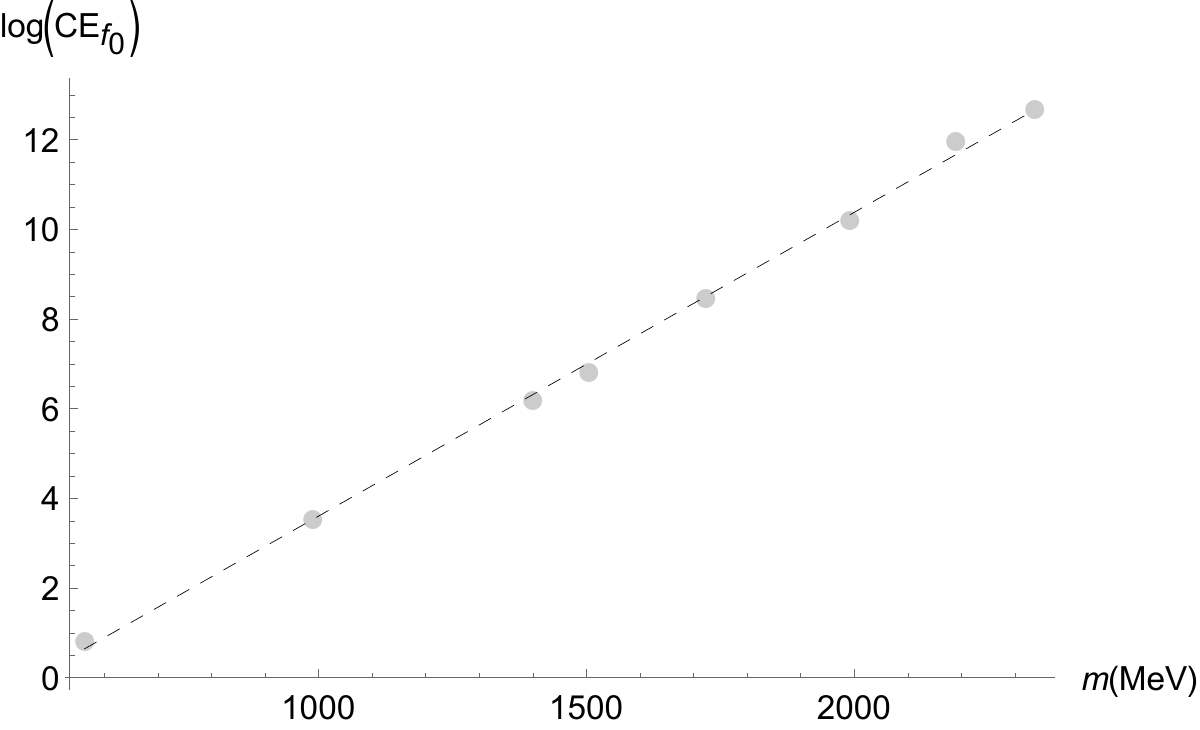}
\caption{Logarithm of the configurational entropy of the $\rho$ (blue), $a_1$, (orange), and  $f_0$ (green) mesons families, as a function of their mass spectra. 
The respective informational entropic Regge trajectories are also plotted.}
\label{f6}
\end{figure}

The meson spectra was computed for the $\Upphi_1(z)$ and $\Upphi_2(z)$ dilaton backgrounds in Ref. \cite{Li:2013oda}. Replacing Eqs. (\ref{quadraticd}) -- (\ref{tanh}),
and the scalar field $\upxi(z)$ in Eqs. (\ref{equacoes}) -- (\ref{eomall3}), one can derive 
 the  ${\rm A}(z)$ warp factor.  
To obtain the $S$ scalar mesons mass spectra, the following action was employed in Ref. \cite{Li:2013oda} 
\begin{eqnarray}\label{lagras}
\! S_{s}\!&\!\!=\!\!& \!\kappa\!\int 
 e^{-\Upphi}\sqrt{-g}\left(\partial_mS\partial^mS\!+\!2S^2\Upphi^2\right)\,\!d^5x,
\end{eqnarray}
where $\partial_mS\partial^mS=\partial_zS\partial^zS\!+\!\partial_\mu S\partial^\mu S$, and $\kappa=-2\frac{N_f}{N_c\ell^3}$. The EOM for the $S$ scalar is given by Eq. \eqref{eomf0}, however with  the Schr\"odinger potential
\begin{eqnarray}\label{eps}
V_{\tt s}(z)&=&\frac14{(\Upphi'-3{\rm A}')^2}+\frac32({3{\rm A}''-\Upphi''})
\nonumber \\
&&  -\left(2{\rm A}'\!-\!\frac{\Upphi'}{2}\!+\!\frac{\Upphi'}{2(1\!+\!\Upphi)}\right)(3{\rm A}'\!-\!\Upphi')\nonumber\\&&\!+\!\left({\rm A}'\!-\!\frac32\Upphi'
\!-\!\frac{\Upphi'}{2(1\!+\!\Upphi)}\right)\!\log(2{\upxi'})'\!+\!\frac{\upxi'''}{\upxi'},
\end{eqnarray}
instead. 
The mass spectra for the $f_0$ scalar meson family is then forthwith obtained, being listed in the second and third rows  of Table \ref{scalarmasses}, respectively for the quadratic (Eq. \eqref{quadraticd}) and the deformed (Eq. \eqref{tanh}) dilaton fields. It is accomplished by solving Eq. \eqref{eomf0}
with the potential in Eq. (\ref{eps}), using boundary condition $\lim_{z\rightarrow\infty} {\tt s}_n'(z)=0$ and ${\tt s}_n(0)=0$, with 
 parameters $m_q\approx9 ~{\rm MeV}$ and $ \mu\approx429~ {\rm MeV}$.  \textcolor{black}{The first column replicates the mass spectra in the PDG 2018 for   
$f_0(500)$, $f_0(980)$, $f_0(1370)$, $f_0(1500)$, $f_0(1710)$, as well as for the $f_0(2020)$, $f_0(2200)$ and $ f_0(2330)$ scalar mesonic states, that are still left out the summary table in PDG (few events registered \cite{pdg1}).}
\begin{table}[h]
\begin{center}-------------- $f_0$ scalar mesons mass spectra ---------------\medbreak
\begin{tabular}{||c|c||c|c||}
\hline\hline
        $n$ & ~Exper. (MeV)        & mass$_{\Upphi_1(z)}$ (MeV)&  mass$_{\Upphi_2(z)}$ (MeV)   \\\hline\hline
\hline
         \textcolor{black}{1*} & $563^{+58}_{-69}$    &420.9     &186.9              \\\hline
         \textcolor{black}{2*}& $990 \pm 20$           &1042.6      &1077.8             \\\hline
         \textcolor{black}{3*} & $1400 \pm 40$         &1369.5                &1434.0   \\\hline
         \textcolor{black}{4*} & $1504 \pm 6$           &1625.0                 &1684.5   \\\hline
         \textcolor{black}{5*} & $1723^{+6}_{-5}$           &1842.4      &1889.7            \\\hline
        6 & $1992 \pm 16$          &2035.7       &2067.4             \\\hline
        7 & $2189 \pm 13$           &2211.9   &2233.8                 \\\hline
        8 &  $2337 \pm 14$         &2374.8   &2391.8                  \\
\hline\hline
\end{tabular}
\caption{The experimental and predicted mass spectra for $f_0$ scalar
mesons, in both the quadratic (\ref{quadraticd})  (second column) and the deformed  (\ref{tanh})  (third column) dilaton profiles. Respectively along the rows, for  the $f_0(500)$, $f_0(980)$, $f_0(1370)$, $f_0(1500)$, $f_0(1710)$, $f_0(2020)$, $f_0(2200)$ and $ f_0(2330)$ mesons. \textcolor{black}{The modes indicated by asterisk are confirmed
states in PDG, whereas the other ones have not been experimentally confirmed states yet \cite{pdg1}.}} \label{scalarmasses}
\end{center}
\end{table}
For deriving the mass spectra for the $\rho$ meson family, for $\upkappa=\frac{N_f}{2g_5^2N_c \ell^3}$, the action 
\begin{eqnarray}
 S_{V}=-\upkappa\int \,e^{5{\rm A}-\Upphi}\,g_{\mu\nu}\partial_m V^\intercal_\mu \partial^mV^{\intercal\nu}\,d^5x,\label{lagrav}
\end{eqnarray}
is employed, where $V^\intercal_m$ denotes the transverse
components in Eq. (\ref{v12}). 
 The normalizable solutions, ${\tt v}_n$, of the associated EOMs (\ref{eomrho}) and (\ref{v10}), are obtained as Kaluza--Klein modes for discrete values of the  4d momentum $q^2=m^2_n$.
The boundary fields ${\tt v}_n$ (cf.  Eq. (\ref{eomrho})) play the role of external sources, coupled to the  QCD current densities. The $\rho$ meson family mass spectra can be then obtained from Eq. (\ref{eomrho}) 
with boundary condition ${\tt v}_n(0)=0$, $\lim_{z\rightarrow\infty} {\tt v}_n'(z)=0$. 
The mass spectra is shown in  the second and third rows  of Table \ref{vmassa}, respectively for quadratic (Eq. \eqref{quadraticd}) and deformed (Eq. \eqref{tanh}) dilaton fields \cite{Li:2013oda}. The first column in Table \ref{vmassa} corresponds to the experimental data in Fig. \ref{f1}. 
\begin{table}[h]
\begin{center}-------------- $\rho$ vector mesons mass spectra -----------------\medbreak
\begin{tabular}{||c||c|c|c||cccc}
\hline\hline
    $n$ & ~Exp. (MeV)        & mass$_{\Upphi_1(z)}$ (MeV)&  mass$_{\Upphi_2(z)}$ (MeV)   \\\hline\hline
        \textcolor{black}{1*} & $775.26 \pm 0.25$         &727.8      & 753.9               \\\hline
        2 & $1350^{+40}_{-50}$         &1134.6               &1133.8                       \\\hline
        \textcolor{black}{3*} & $1465 \pm 25$         &1426.0    &1430.0            \\\hline
        4& $1570\pm 98$         &1534.1& 1537.9\\\hline
        \textcolor{black}{5*} & $1720 \pm 20$         &1664.5                &1667.9    \\\hline
        6 & $1909 \pm 30$         &1873.6      &1875.4            \\\hline
        7 & $2149 \pm 17$         &2061.9     &2063.7              \\\hline
        8 & $2265 \pm 40$         &2233.6     &2234.6              \\
\hline\hline
\end{tabular}
\caption{The experimental \cite{pdg1} and predicted mass spectra for the $\rho$ vector meson family, in both the quadratic (Eq. (\ref{quadraticd})) and the deformed (Eq. (\ref{tanh})) dilaton profiles. Respectively along the rows, for the $\rho(770)$, $\rho'(1450)$, $\rho(1450)$, $\rho(1570)$, $\rho(1700),$ $\rho(1900)$, $\rho(2150)$ and  $\rho(2270)$ mesons. \textcolor{black}{The modes indicated by asterisk are experimentally confirmed
states, whereas the other ones are omitted from the summary table in PDG \cite{pdg1}.}} \label{vmassa}
\end{center}
\end{table}
Similarly to the $\rho$ mesons, the ${\tt a}_n$ axial vector mesonic excitations (cf. Eq. (\ref{eoma1})) describe the 
$a_1$ axial vector meson family, whose  mass spectra  can
be obtained from the modes of the axial gauge field (\ref{a12}) in the bulk.
The quadratic terms in the transverse component of 
the axial vector $\mathring{A}^m$ (\ref{a12}), denoted by $\mathring{A}^\intercal$,  are used to construct the action 
\begin{eqnarray}
\!\!\!\!\!\!\! S_{\mathring{A}}\!=\!\upkappa\!\!\int e^{5{\rm A}-\Upphi}\!\left(\partial_m \mathring{A}^\intercal_\mu \partial^m\mathring{A}^{\intercal\mu}\!+\!\frac{4g_5^2\upxi^2}{\ell^2} \mathring{A}^{\intercal}_{\mu} \mathring{A}^{\intercal\mu}\!\right)\,d^5x,\label{lagraa}
\end{eqnarray} 
where $g_5^2={4\pi^2\,N_f}/{N_c}$.
The bulk effective mass in (\ref{lagraa}) is generated by the Higgs mechanism, with the $\mathfrak{X}$ scalar field encoding the chiral symmetry
breaking \cite{EKSS2005}.  The EOMs that drive the axial vector mesonic states can be written as Eqs. (\ref{eoma1}) -- (\ref{v11}),  with boundary conditions $\lim_{z\rightarrow\infty} {\tt a}_n'(z)=0$ and ${\tt a}_n(0)=0$. The mass spectra of the $a_1$ axial vector meson family   is shown in  the second and third rows  of Table \ref{avectormasses}, respectively for  quadratic (Eq. \eqref{quadraticd}) and deformed (Eq. \eqref{tanh}) dilaton fields. The first column in Table \ref{avectormasses} represents the experimental data \cite{pdg1}. \textcolor{black}{For $a_1(1260)$, $a_1(1420)$, $a_1(1640)$ axial mesons experimentally confirmed, as well as for $a_1(1930)$, $a_1(2095)$ and $a_1(2270)$, the axial vector mesonic states  omitted from the summary table in PDG \cite{pdg1} (few events registered). }
\begin{table}[H]
\begin{center}------------------ $a_1$ mesons mass spectra --------------------\medbreak
\begin{tabular}{||c||c|c|c||cccc}
\hline\hline
   $n$ & ~Exp. (MeV)        & mass$_{\Upphi_1(z)}$ (MeV)&  mass$_{\Upphi_2(z)}$ (MeV)   \\\hline\hline
        \textcolor{black}{1*} & $1255^{+13}_{-23}$           &1064.9    &   1117.5      \\\hline
        \textcolor{black}{2*}& $1414^{+15}_{-13}$      & 1364.4&1624.1\\\hline
        \textcolor{black}{3*} & $1654 \pm 19$           &1561.7       & 1623.0     \\\hline
        4 & $1930^{+30}_{-70}$      &1845.8    &1878.6         \\\hline
        5 & $2096 \pm 138$          &2059.0 &2082.5           \\\hline
        6 & $2265 \pm 50$      &2242.0  &2262.7            \\\hline
\hline\hline
\end{tabular}
\caption{The experimental \cite{pdg1} and predicted mass spectra for $a_1$  axial
vector mesons, in both the quadratic (Eq. (\ref{quadraticd})) and the deformed (Eq. (\ref{tanh})) dilaton profiles. Respectively along the rows, for $a_1(1260)$, $a_1(1420)$, $a_1(1640)$, $a_1(1930)$, $a_1(2095)$ and $a_1(2270)$ mesons. \textcolor{black}{The modes indicated by asterisk are confirmed
states in PDG \cite{pdg1}.}} \label{avectormasses}
\end{center}
\end{table}
Now, in order to compute the CE for the $f_0$, the $\rho$, and the $a_1$ mesons  families, and  one first considers a localized, Lebesgue-integrable $\epsilon(z)$  energy density, associated to each meson family.
In general, given an arbitrary Lagrangian, $\mathfrak{L}$, the energy-momentum tensor reads 
 \begin{equation}
 \!\!\!\!\!\!\!\!T^{mn}\!=\!  \frac{2}{\sqrt{ - g }}\!\! \left( \frac{\partial (\sqrt{-g} \mathfrak{L})}{\partial g_{mn} }\!-\!\partial_{ x^q }  \frac{\partial (\sqrt{-g}  \mathfrak{L})}{\partial \left(\!\frac{\partial g_{mn} }{\partial x^q}\!\right) }
%  \!+\!\mathcal{T}^{mn}\!
  \right).
  \label{em1}
 \end{equation} 
 \noindent  
 %where either $\mathcal{T}^{mn}\!=\!\frac{\partial\mathfrak{L}}{\partial(\partial_m\Upphi)}\partial^n\Upphi\!+\!\frac{\partial\mathfrak{L}}{\partial(\partial_m S)}\partial^nS\!-\!\delta^{mn}\mathfrak{L}$, for the scalar Lagrangian (\ref{lagras}), or 
 %$\mathcal{T}^{mn}=\frac{\partial\mathfrak{L}}{\partial(\partial_m B_q)}\partial^n B_q-\delta^{mn}\mathfrak{L}$, where $B_m$ is either the vector or the axial vector field, respectively in Eqs. (\ref{lagrav}) and (\ref{lagraa}). 
 The $\epsilon(z)$ energy density corresponds to the $T_{00}(z) $ component of (\ref{em1}), respectively for the $f_0$ meson family (\ref{lagras}), for the $\rho$ meson family (\ref{lagrav}) and for the $a_1$ meson family (\ref{lagraa}). 
The Fourier transform 
$\epsilon(k) = \int_\mathbb{R}\epsilon(z)e^{-ik\cdot z}\,dz,$ with respect to the $z$ dimension that defines the energy scale in AdS/QCD, is then employed to define the modal fraction    
\cite{Gleiser:2012tu,Sowinski:2015cfa}
\begin{eqnarray}
\upepsilon(k) = \frac{|\epsilon(k)|^{2}}{ \int_{\mathbb{R}}  |\epsilon(k)|^{2}dk}.\label{modalf}
\end{eqnarray} It is a correlation probability distribution that quantifies how much a $k$ wave  mode contributes to the power spectrum, associated to the energy density.
The CE, then, measures the information content of the spatial profile that characterizes the energy density, $\epsilon(z)$, with respect to the Fourier wave modes. The CE 
 is defined by \cite{Gleiser:2012tu,Sowinski:2015cfa}
\begin{eqnarray}
S[\epsilon] = - \int_{\mathbb{R}}{\upepsilon_\diamond}(k)\log {\upepsilon_\diamond}(k)\, dk\,,
\label{confige}
\end{eqnarray}
for $\upepsilon_\diamond(k)=\upepsilon(k)/\upepsilon_{\rm max}(k)$.
After numerical calculations, the CE is obtained as a function of the $n$ excitation number, $1\leq n \leq 8$, for the $a_1$ axial vector, the $\rho$ vector, and the $f_0$ scalar mesons. The results are listed in Tables \ref{CEmesons} and \ref{CEmesonstanh}, respectively for the quadratic and deformed dilatonic field backgrounds. 
\begin{table}[h]
\begin{center}%---------------------------------------------------------------------\\
------------------------ $\Phi_1(z)=\mu_{\rm G}^2z^2$ ------------------------\medbreak
\begin{tabular}{||cc||c|c|c||c||c||}
\hline\hline
  &   $n$ & $\rho$ mesons~CE & $a_1$ mesons~CE &$f_0$ mesons~CE \\ \hline\hline
     &  \, 1 \,&\, $32.8$   \,&\,  338.2   \,&\, 3.26\\\hline
     \,&\,   2 \,&\, $307.2$ \,&\, $1.612\times 10^3$  \,&\,34.81   \\\hline
     \,&\,   3 \,&\, $2.108\times 10^3$  \,&\,$1.027\times 10^4$    \,&\,222.4  \\\hline
     \,&   4\, &\, $9.832\times 10^3$  \,&\,$5.804\times 10^4$   \,&\,913.1  \\\hline
     \,&\,   5 \,&\, $4.902\times 10^4$   \,&\,$2.421\times 10^5$   \,&\, 4.723$\times 10^3$ \\\hline
     \,&\,   6 \,&\, $2.092\times 10^5$   \,&\,$1.614\times 10^6$  \,&\,  2.688$\times 10^4$\\\hline
     \,&\,  7 \,&\, \,\,$7.721\times 10^5$   \,&\,$1.057\times 10^7$  \,&\,1.571$\times 10^5$    \\\hline
     \,&\,8\,&$\,3.423\times 10^6$\,&\,$2.762\times 10^7$\,&\,$5.241\times 10^5$\\
\hline\hline
\end{tabular}
\caption{The CE for the $\rho$ vector, the $a_1$ axial vector, and the $f_0$ scalar  mesons families,  in the $\Upphi_1(z)=\mu_{\rm G}^2z^2$ quadratic dilaton soft wall model. }
\label{CEmesons}\end{center}
\end{table}\begin{table}[h]
\begin{center}
%-----------------------------------------------------------------------\\
------------ $
 \Phi_2(z)=\mu_{\rm G}^2z^2\tanh(\mu_{{\rm G}^2}^4z^2/\mu_{\rm G}^2)$ -------------- \medbreak
\begin{tabular}{||cc||c|c|c||c||c||}
\hline\hline
    &   $n$ & $\rho$ mesons~CE & $a_1$ mesons~CE &$f_0$ mesons~CE \\ \hline\hline
     &  \, 1 \,&\, $28.05$   \,&\,  587.51   \,&\, 4.28\\\hline
     \,&\,   2 \,&\, $547.49$ \,&\, $2.621\times 10^3$  \,&\,44.32   \\\hline
     \,&\,   3 \,&\, $2.223\times 10^3$  \,&\,$1.518\times 10^4$    \,&\,288.31  \\\hline
     \,&   4\, &\, $1.016\times 10^4$  \,&\,$7.142\times 10^4$   \,&\,1.528$\times 10^3$  \\\hline
     \,&\,   5 \,&\, $5.966\times 10^4$   \,&\,$2.982\times 10^5$   \,&\, 8.984$\times 10^3$ \\\hline
     \,&\,   6 \,&\, $3.417\times 10^5$   \,&\,$1.728\times 10^6$  \,&\,  5.347$\times 10^4$\\\hline
     \,&\,  7 \,&\, \,\,$2.017\times 10^5$   \,&\,$7.135\times 10^6$  \,&\,2.491$\times 10^5$    \\\hline
     \,&\,8\,&$\,5.822\times 10^6$\,&\,$3.231\times 10^7$\,&\,$1.167\times 10^6$\\
\hline\hline
\end{tabular}
\caption{The CE for the $\rho$ vector,  the $a_1$ axial vector, and the $f_0$ scalar mesons families, in the $\Upphi_2(z)=z^2\tanh\left(\frac{\mu_{{\rm G}^2}^4z^2}{\mu_{\rm G}^2}\right)$ deformed dilaton soft wall model. }
\label{CEmesonstanh}
\end{center}
\end{table}
In both Tables \ref{CEmesons} and \ref{CEmesonstanh}, respectively along the rows, the first column is depicted for the $\rho$ vector meson family, identifying the ${\tt v}_n$ eigenfunctions  in Eq. (\ref{eomrho}) as ${\tt v}_1=\rho(770)$, ${\tt v}_2=\rho'(1450)$, ${\tt v}_3=\rho(1450)$, ${\tt v}_4=\rho(1570)$, ${\tt v}_5=\rho(1700),$ ${\tt v}_6=\rho(1900)$, ${\tt v}_7=\rho(2150)$ and  ${\tt v}_8=\rho(2270)$. The second columns in Tables \ref{CEmesons} and  \ref{CEmesonstanh} are composed by the ${\tt a}_n$ eigenfunctions in Eq. (\ref{eoma1}), describing the $a_1$ axial vector meson family by the  identification ${\tt a}_1=a_1(1260)$, ${\tt a}_2=a_1(1420)$, ${\tt a}_3=a_1(1640)$, ${\tt a}_4=a_1(1930)$, ${\tt a}_5=a_1(2095)$ and ${\tt a}_6=a_1(2270)$. 
The mesonic excitations ${\tt a}_7$ and ${\tt a}_8$ are solutions of the EOM (\ref{eomf0}), for $n=7$ and $n=8$, not detected yet. Besides, the third columns  in Tables \ref{CEmesons} and \ref{CEmesonstanh} show the $f_0$ scalar meson family, described by the ${\tt s}_n$ wave eigenfunctions of Eq. (\ref{eomf0}). They read  ${\tt s}_1=f_0(500)$, ${\tt s}_2=f_0(980)$, ${\tt s}_3=f_0(1370)$, ${\tt s}_4=f_0(1500)$, ${\tt s}_5=f_0(1710)$, ${\tt s}_6=f_0(2020)$, ${\tt s}_7=f_0(2200)$ and ${\tt s}_8= f_0(2330)$. 
Firstly analyzing the quadratic dilatonic potential in Eq. (\ref{quadraticd}), let one takes the 
CE for the $\rho$, $a_1$, and $f_0$ meson families listed in Table \ref{CEmesons}. Computing the logarithm of the CE for each $n$ mode for the three mesons families  leads to the result depicted in Fig. \ref{f2}, whose numerical interpolation provides the first type of informational entropic Regge trajectories. For the $\rho$ vector, the $a_1$ axial vector, and the $f_0$ scalar meson families, respectively, the informational entropic Regge trajectories are the dotted lines in Fig. \ref{f2}.
\noindent Their explicit expressions are, respectively, 
\begin{eqnarray}
 \log({\rm CE}_\rho(n)) &=& 1.6123\,n + 2.4594,\\
 \log({\rm CE}_{a_1}(n)) &=& 1.6632 \,n + 4.1923,\\
 \log({\rm CE}_{f_0}(n)) &=& 1.6907\,n + 0.0150,
   \end{eqnarray} within respectively $0.9\%$, $1.2\%$, and $\sim1.7\%$ standard deviations.
Now, with the computed CE for the $\rho$, $a_1$, and $f_0$ meson families in the $\Phi_2$ dilaton background (\ref{tanh}), listed in Table \ref{CEmesonstanh}, one can also calculate the logarithm of the CE, for each $n$ excitation mode. Fig. \ref{f3} shows the corresponding results for each meson family, wherein linear regression yields the second type of informational entropic Regge trajectories. 
\noindent The explicit expressions for each informational entropic Regge trajectory is numerically obtained by interpolation of the data in Table \ref{CEmesonstanh}. The obtained linear regressions, respectively for the $a_1$ axial vector, the $\rho$ vector, and the $f_0$ scalar mesons families, read 
\textcolor{black}{\begin{eqnarray}\label{itp1}
 \log({\rm CE}_\rho(n)) &=& 1.8497\,n + 1.7763, \\
 \log({\rm CE}_{a_1}(n)) &=& 1.5623 \,n + 4.9425,\label{itp2}\\
\label{itp3}
 \log({\rm CE}_{f_0}(n)) &=& 1.7624\,n + 0.2112,
   \end{eqnarray} within $\sim0.6\%$, $\sim1.4\%$, and $\sim1.7\%$ standard deviations, respectively.}
For the $\rho$, $a_1$, and $f_0$ mesons families, one can hence realize a  scaling law, relating the logarithm of the CE and the $n$ meson excitation modes. 
Figs. \ref{f2} and \ref{f3}
show that there are informational entropic Regge trajectories, implementing a  relation between the logarithm of the CE and the $n$ excitation number, for both the quadratic and deformed dilatonic potentials.  %The explicit scaling law, for both the $\Upphi_1$ and $\Upphi_2$ potentials  is given in Table \ref{law}. 
%\begin{table}[h]   \label{law}
%\begin{center}
%\begin{tabular}{||c||c||c|c|c||c||c||}
%\hline\hline
 % & ${\rm CE}_\rho(n)$ &  ${\rm CE}_{a_1}(n)$ &${\rm CE}_{f_0}(n)$ \\ \hline\hline
% $\Upphi_1(z)$& $14.65\,e^{1.601\,n}$&$141.74\,e^{1.458 \,n}$&$1.736\,e^{1.5957\,n}$\\\hline
% $\Upphi_2(z)$&$14.64\,e^{1.640\,n}$&$141.75\,e^{1.572 \,n}$&$1.223\,e^{1.7936\,n}$\\\hline
%   \hline
%\end{tabular}
%\caption{The CE scaling law as a function of the $n$ excitation number, for $\rho$ vector mesons, $a_1$ axial vector mesons, $f_0$
%scalar mesons, both  in the $\Upphi_1(z)=\mu_{\rm G}^2z^2$ quadratic dilaton soft wall model, 
%with $m_q\approx 9 {\rm MeV},\mu\approx 429~ {\rm MeV}$. }
%\label{CEmesons}
%\end{center}
%\end{table}
%
The original Regge trajectories in the soft wall AdS/QCD regard the relation $m_n\sim n$, for the light-flavor meson mass spectra. One can then emulate them in the information entropic context. In fact, one can calculate the logarithm of the CE for each meson family, as a function of the meson mass spectra, experimentally detected.
In what follows, the $\Upphi_2(z)=z^2\tanh\left(\frac{\mu_{{\rm G}^2}^4z^2}{\mu_{\rm G}^2}\right)$ deformed dilaton in the soft wall AdS/QCD model shall be employed, as it better describes the meson mass spectra.
The results are plotted in Fig. \ref{f6}.
\begin{figure}[H]
\centering
\includegraphics[width=7.2cm]{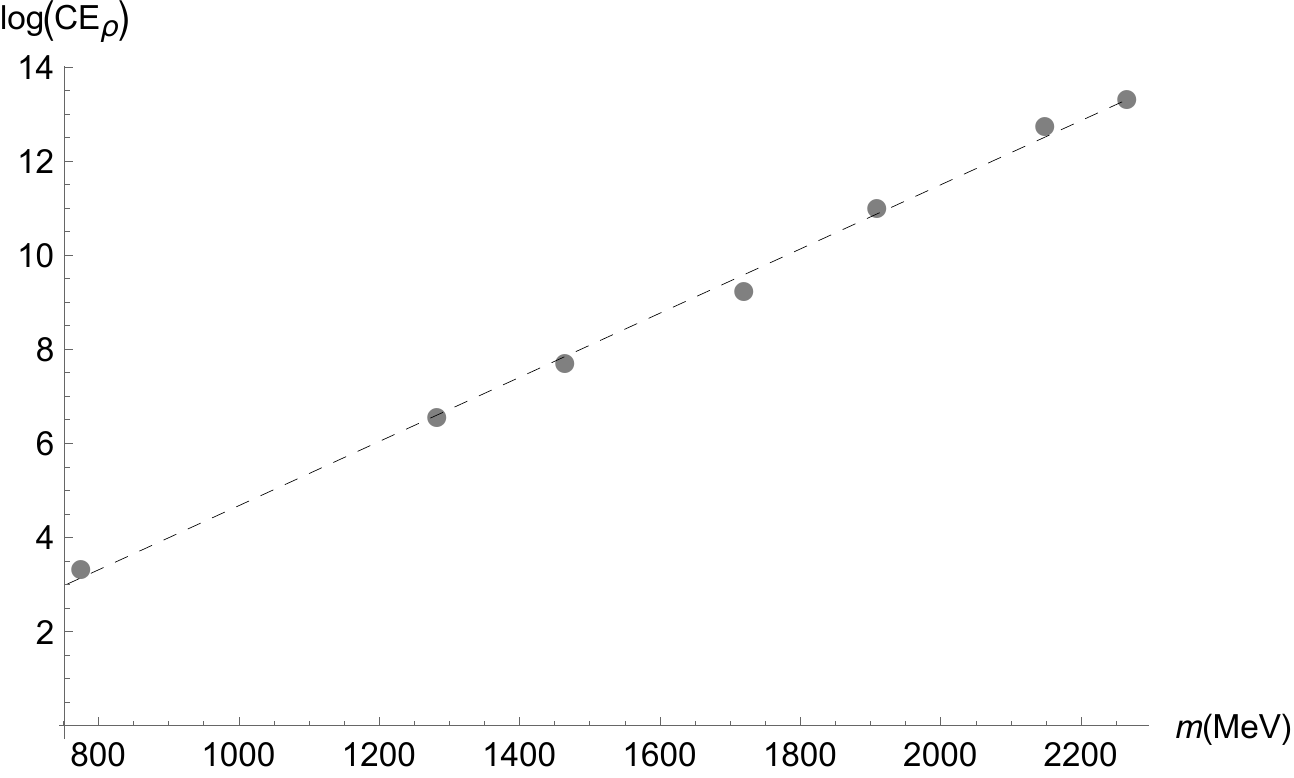}
\label{f4}\end{figure}
\noindent The informational entropic Regge trajectories, as a function of the meson mass, $m$ (MeV), are respectively listed as follows:
\textcolor{black}{\begin{eqnarray}\label{itq1}
 \log({\rm CE}_\rho(m)) &=& 0.0067\,m -1.9651, \\
 \log({\rm CE}_{a_1}(m)) &=& 0.0069 \,m -2.2184,\label{itq2}\\
\label{itq3}
 \log({\rm CE}_{f_0}(m)) &=& 0.0062\,m -2.2013,
   \end{eqnarray} within $\sim2.6\%$, $\sim2.1\%$, and $\sim2.5\%$ standard deviations, respectively.}
The informational entropic Regge trajectories in Eqs. (\ref{itq1}) -- (\ref{itq3}) bring 
another very interesting aspect of the CE that underlies the meson families. Instead of computing
the meson family mass spectra, solving Eqs. (\ref{eomf0}) -- (\ref{eoma1}), one can extrapolate the interpolation lines (\ref{itp1}) -- (\ref{itp3}) to compute the CE for the meson families, at  least for the $n$ mesonic excitations such that $n> 8$, for the $\rho$ and $f_0$ families, and such that $n>6$, for the $a_1$ family.  

In the following discussion of the informational entropic Regge trajectories (\ref{itp1}) -- (\ref{itp3}), the notation $m_{\alpha,n}$ means the mass of the $n^{\rm th}$ meson in some $\alpha$-meson family ($\alpha=\rho, a_1, f_0$),  corresponding to the element in the respective meson family with $n$ excitation number. Considering the experimental of the $\rho$ vector meson family, although there is no mesonic excitation ${\tt v}_n$ in the $\rho$ meson family
higher than $n=8$, experimentally detected, 
the informational entropic Regge trajectory in Fig. \ref{f6}
deploys a reliable method for predicting the mass of the ${\tt v}_n$ vector meson states, for $n\geq 9$.
Although the explicit calculation by Eq. (\ref{eomrho}), with potential (\ref{v10}), is already known to produce the mass spectra of the $\rho$ vector meson family, as a function of the $n$ excitation number, here the masses of the ${\tt v}_9$ and ${\tt v}_{10}$ elements in the $\rho$ meson mass family can be then inferred. \textcolor{black}{In fact, for $n=9$, Eq. (\ref{itp1}) yields 
$\log({\rm CE}_\rho) = 18.423$.  Replacing this value in the informational entropic Regge trajectory (\ref{itq1}), one obtains the mass $m_{\rho,9}=2878$ MeV, for the ${\tt v}_9$ mesonic state in the $\rho$ meson family. The standard deviations for Eqs. (\ref{itp1}) -- (\ref{itq1}) gives the reliable range $2813\; {\rm MeV}\lesssim m_{\rho,9}\lesssim 2945\;{\rm MeV}$, for the ${\tt v}_9$ vector meson state. Similarly, the ${\tt v}_{10}$ vector meson excitation has mass $m_{\rho,10}=3098$ MeV. Considering the standard deviations for Eqs. (\ref{itp1}) -- (\ref{itq1}), 
the range $3042\lesssim m_{\rho,10}\lesssim 3153$ MeV is a reliable one. One can further extrapolate the mass spectra for the ${\tt v}_n$ vector mesonic excitations, for $n\geq 11$, however the standard deviations are larger, the higher the $n$ excitation number is. 
Analogously, the $a_1$ axial vector meson family can be analyzed.
The masses of the ${\tt a_7}$ and ${\tt a}_{8}$ can be then inferred, using the 
mass spectra of the experimentally detected mesons in this family, using Eq. (\ref{itp2}) --  (\ref{itq2}). In fact, for $n=7$, Eq. (\ref{itp2}) yields 
$\log({\rm CE}_{a_1}) = 15.878$.  Replacing this value in the informational entropic Regge trajectory (\ref{itq1}), one obtains the mass $m_{a_1,7}=2567$ MeV, for the ${\tt a}_7$ mesonic state in the $a_1$ meson family. The standard deviations for Eqs. (\ref{itp2}) and  (\ref{itq2}) gives the reliable range $2491\; {\rm MeV}\lesssim m_{a_1,7}\lesssim 2637\;{\rm MeV}$, for the ${\tt a}_7$ axial vector meson state. Similarly, the ${\tt a}_8$ vector meson excitation has mass  $m_{a_1,8}=2782$ MeV.  The standard deviations for Eqs. (\ref{itp1}) and  (\ref{itq1}) yield 
the range $2708\lesssim m_{a_1,8}\lesssim 2869$ MeV. One can further infer the mass spectra of the ${\tt a}_n$ vector mesonic excitations, for $n\geq 9$. 
%It is worth to emphasize that as this scaling rule was based on the $n$ excitation numbers in the range $1\leq n \leq 6$, nothing may make us to assert that 
%this scaling rule holds for $n\geq 9$. However, one can try to 
%use the scaling rules in Table \ref{***} to compute the CE, and then 
%
Finally, the masses of the next generation of $f_0$ scalar mesons, can be predicted.
Employing Eq. (\ref{itp3}), for $n=9$ implies that  
$\log({\rm CE}_{f_0}) = 16.073$. Now, one can substitute this value into Eq. (\ref{itq3}),  yielding the mass of the ${\tt s}_9$ element in the $f_0$ meson family, $m_{f_0,9}=2905$ MeV. The standard deviations for Eqs. (\ref{itp3}) -- (\ref{itq3}) gives the reliable range $2806\; {\rm MeV}\lesssim m_{f_0, 9}\lesssim 2996\;{\rm MeV}$, for the ${\tt s}_9$ scalar  mesonic excitation. Besides, similar calculations yield  the ${\tt s}_{10}$ scalar meson excitation  mass, $m_{f_0,10}=3189$ MeV. The standard deviations related to Eqs. (\ref{itp1}) and (\ref{itq1}) imply the range $3094\lesssim m_{\rho,10}\lesssim 3301$ MeV, for the ${\tt s}_{10}$ scalar meson. One can further extrapolate the mass spectra of the ${\tt s}_n$ scalar mesonic excitations, for $n\geq 11$, however the standard deviations increase, the higher the $n$ excitation number is.}

\vspace*{-0.3cm}
\section{Concluding remarks and perspectives}
\label{iv}

The CE was computed for $a_1$ axial vector, the $\rho$ vector, and the $f_0$ scalar mesons families, for two dilatonic backgrounds, in a graviton-dilaton-gluon background. Two types of informational entropic Regge trajectories were derived, for each meson family. 
The first one consists of the CE in terms of the meson $n$ excitation number, described 
by Eqs. (\ref{itp1}) -- (\ref{itp3}) and illustrated in Figs. \ref{f2} and \ref{f3}, for both quadratic and deformed dilatonic profiles. The second  type of informational entropic Regge trajectories
relates the logarithm of the CE to the experimental  mass spectra of the mesons families, in Eqs. (\ref{itq1}) -- (\ref{itq3}), respectively shown in the plots of Fig. \ref{f6}. Consequently, the mesons families mass spectra were extrapolated  
from these informational entropic Regge trajectories. A range for the mass spectra  
of mesons with higher $n$ excitation numbers, in each meson family, was then estimated with good accuracy. The two first elements of the next generation, in each meson family, were studied and discussed. 

The prediction of the meson mass spectra through 
Eqs. (\ref{eomf0}) -- (\ref{eoma1}), although also taking experimental parameters
to fit the meson mass spectra, is a theoretical prediction, that already 
matches experimental data, as shown in Ref. \cite{Li:2013oda} and illustrated in Tables \ref{scalarmasses} -- \ref{avectormasses}. On the hand, the very essence of procedure throughout Sect. \ref{ivi} estimates the mass spectra of the next generation of mesons  by the informational entropic Regge trajectories, based on the mass spectra of the already detected mesons. Indeed, the first types (\ref{itp1}) -- (\ref{itp3}) of informational entropic Regge trajectories express  the CE, once the 
$n$ excitation number is fixed. With the obtained value of the CE, Eqs. (\ref{itq1}) -- (\ref{itq3}) 
then determine the values of the masses of the next generation of mesonic states, in each $\rho$, $a_1$, and $f_0$ family. Since the informational entropic Regge trajectories are determined by the experimental meson mass spectra, then this procedure can determine at least the two next 
elements in each meson family, with good accuracy. The eventual detection of these new mesonic states shall contribute with more experimental points in the plots of Fig. \ref{f6},
improving the fitting of Eqs. (\ref{itq1}) -- (\ref{itq3}). 

Pseudo-scalar mesons can be further regarded, whose CE may be also computed. 
However, their Lagrangian involves a pseudo-scalar field that is coupled to a
$\varphi$ scalar field that defines the parallel axial vector (complementary to the transverse to the $\mathring{A}^\intercal_\mu$ field  in (\ref{lagras}))  as $\partial_\mu \varphi$. Hence, the derived coupled system of EOMs involve awkward 
Schr\"odinger-like potentials, turning the CE computation a hard task, unsolved up to now.
Besides, as the soft wall AdS/QCD model corresponds to the $D_3$ - $D_q$, system, 
extensions involving $D_p$ - $D_q$ models \cite{Huang:2007fv} may be also accomplishable, although it is far beyond the scope here assumed. 
Finite temperature effects in the soft wall AdS/QCD may be also implemented, as
whose initial results using the CE apparatus were introduced in Ref. \cite{Braga:2016oem} for quarkonia.

\paragraph*{Acknowledgments:}   The work of AEB is supported by the Brazilian Agencies FAPESP (grant 2018/03960-9) and CNPq (grant 300831/2016-1). RdR~is grateful to FAPESP (Grant No.  2017/18897-8) and to the National Council for Scientific and Technological Development  -- CNPq (Grant No. 303293/2015-2), for partial financial support.

%%%%% CLEAR DOUBLE PAGE!
\newpage{\pagestyle{empty}\cleardoublepage}


\begin{thebibliography}{9}

\bibitem{shannon} C. E. Shannon, Bell Syst. Tech. J. {\bf 27} (1948) 379; 623.

\bibitem{Gleiser:2011di} M. Gleiser and N. Stamatopoulos, Phys.\ Lett.\ B {\bf 713} (2012) 304 [{arXiv:1111.5597 [hep-th]}].

\bibitem{Gleiser:2012tu} M. Gleiser and N. Stamatopoulos, Phys.\ Rev.\ D {\bf 86} (2012) 045004 [{arXiv:1205.3061 [hep-th]}].

\bibitem{Gleiser:2018kbq} M.~Gleiser, M.~Stephens and D.~Sowinski,
  %``Configurational entropy as a lifetime predictor and pattern discriminator for oscillons,''
  Phys.\ Rev.\ D {\bf 97}  (2018) 096007 
  [{arXiv:1803.08550 [hep-th]}].

\bibitem{Gleiser:2014ipa} M.~Gleiser and N.~Graham,
  %``Transition To Order After Hilltop Inflation,''
  Phys.\ Rev.\ D {\bf 89} (2014) 083502   [{arXiv:1401.6225 [astro-ph.CO]}].

\bibitem{Sowinski:2015cfa} M. Gleiser and D. Sowinski, Phys.\ Lett.\ B {\bf 747} (2015) 125  [{arXiv:1501.06800 [cond-mat.stat-mech]}].

\bibitem{Bernardini:2016hvx} A.~E.~Bernardini and R.~da Rocha,
  %``Entropic information of dynamical AdS/QCD holographic models,''
  Phys.\ Lett.\ B {\bf 762} (2016) 107  
  %%doi:10.1016/j.physletb.2016.09.023
  [{arXiv:1605.00294 [hep-th]}].


\bibitem{Bernardini:2016qit} A.~E.~Bernardini, N.~R.~F.~Braga and R.~da Rocha,
  %``Configurational entropy of glueball states,''
  Phys.\ Lett.\ B {\bf 765}  (2017) 81  [{arXiv:1609.01258v1 [hep-th]}].

\bibitem{Braga:2017fsb} N.~R.~F.~Braga and R.~da Rocha,
  %``AdS/QCD duality and the quarkonia holographic information entropy,''
  Phys.\ Lett.\ B {\bf 776} (2018) 78 	[{arXiv:1710.07383 [hep-th]}].



\bibitem{Sowinski:2017hdw} D.~Sowinski and M.~Gleiser,
  %``Information Dynamics at a Phase Transition,''
  J.\ Stat.\ Phys.\  {\bf 167} (2017) no.5,  1221
  % doi:10.1007/s10955-017-1762-6
  [{arXiv:1606.09641 [cond-mat.stat-mech]}].

\bibitem{Braga:2016wzx} N.~R.~F.~Braga and R.~da Rocha,
  %``Configurational entropy of anti-de Sitter black holes,''
  Phys.\ Lett.\ B {\bf 767} (2017) 386 [{arXiv:1612.03289 [hep-th]}].

\bibitem{Witten:2018zva} E.~Witten,
  \emph{A Mini-Introduction To Information Theory,} 
  [{arXiv:1805.11965 [hep-th]}].

\bibitem{Natsuume:2014sfa} M.~Natsuume,
  ``AdS/CFT Duality User Guide,''
  Lect.\ Notes Phys.\  {\bf 903} (2015) 1  
  [arXiv:1409.3575 [hep-th]].

\bibitem{Polchinski:2001tt} J.~Polchinski and M.~J.~Strassler,
  %``Hard scattering and gauge/string duality,''
  Phys.\ Rev.\ Lett.\  {\bf 88} (2002) 031601 
  [{arXiv:hep-th/0109174}].

\bibitem{BoschiFilho:2002ta} H.~Boschi-Filho and N.~R.~F.~Braga,
  %``QCD / string holographic mapping and glueball mass spectrum,''
  Eur.\ Phys.\ J.\  C {\bf 32} (2004) 529 
  [{arXiv:hep-th/0209080}].
  %%CITATION = EPHJA,C32,529;%%

\bibitem{Csaki} C. Csaki and M. Reece, JHEP \textbf{05} (2007) 062 [{arXiv:hep-ph/0608266}].

\bibitem{Karch:2006pv} A.~Karch, E.~Katz, D.~T.~Son and M.~A.~Stephanov,
  %``Linear confinement and AdS/QCD,''
  Phys.\ Rev.\ D {\bf 74} (2006) 015005 
 % doi:10.1103/PhysRevD.74.015005
  [{hep-ph/0602229}].

\bibitem{Brodsky:2014yha} S.~J.~Brodsky, G.~F.~de Teramond, H.~G.~Dosch, J.~Erlich,
  %``Light-Front Holographic QCD and Emerging Confinement,''
  Phys.\ Rept.\  {\bf 584} (2015) 1 
 % doi:10.1016/j.physrep.2015.05.001
  [{arXiv:1407.8131 [hep-ph]}].

\bibitem{Li:2013oda} D.~Li and M.~Huang,
  %``Dynamical holographic QCD model for glueball and light meson spectra,''
  JHEP {\bf 1311} (2013) 088 
 % doi:10.1007/JHEP11(2013)088
  [{arXiv:1303.6929 [hep-ph]}].
  %%CITATION = doi:10.1007/JHEP11(2013)088;%%
  
  %%CITATION = PRLTA,88,031601;%%

\bibitem{pdg1} M.~Tanabashi {\it et al.} [ParticleDataGroup],
  %``Review of Particle Physics,''
  Phys.\ Rev.\ D {\bf 98} (2018)  030001.

\bibitem{okubo} S. Okubo, Phys. Lett. {\bf 5} (1963) 165.

\bibitem{Gkk} T.~Gherghetta, J.~I.~Kapusta and T.~M.~Kelley,
%  ``Chiral symmetry breaking in the soft wall AdS/QCD model,''
Phys.\ Rev.\ D {\bf 79} (2009) 076003 [{arXiv:0902.1998 [hep-ph]}].  %%CITATION = ARXIV:0902.1998;%%

\bibitem{rold} L. Da Rold and A. Pomarol, 
%?Chiral Symmetry Breaking from Five Dimensional
%Spaces,? 
Nucl. Phys. B {\bf 721} (2005) 79 [{arXiv:hep-ph/0501218}].

\bibitem{zhang} P. Zhang, 
%?Linear Confinement for Mesons and Nucleons in AdS/QCD?, 
JHEP {\bf 1005}
(2010) 039 [{arXiv:1003.0558}].

\bibitem{sui1} Y.~-Q.~Sui, Y.~-L.~Wu, Z.~-F.~Xie and Y.~-B.~Yang,
  %``Prediction for the Mass Spectra of Resonance Mesons in the soft wall AdS/QCD with a Modified 5d Metric,''
  Phys.\ Rev.\ D {\bf 81} (2010) 014024 [{arXiv:0909.3887 [hep-ph]}].

\bibitem{Colangelo:2011sr}
  P.~Colangelo, F.~Giannuzzi, S.~Nicotri and V.~Tangorra,
  %``Temperature and quark density effects on the chiral condensate: An AdS/QCD study,''
  Eur.\ Phys.\ J.\ C {\bf 72} (2012) 2096
 % doi:10.1140/epjc/s10052-012-2096-9
  [arXiv:1112.4402 [hep-ph]].

\bibitem{hooft} G. 't Hooft, 
 %?A Planar Diagram Theory for Strong Interactions,? 
 Nucl. Phys. B {\bf 72} (1974) 461.

\bibitem{Colangelo:2018mrt}
  P.~Colangelo and F.~Loparco,
  ``Configurational Entropy can disentangle conventional hadrons from exotica,'' 
  [arXiv:1811.05272 [hep-ph]].
  
\bibitem{Barbosa-Cendejas:2018mng} N.~Barbosa-Cendejas, R.~Cartas-Fuentevilla, A.~Herrera-Aguilar, R.~R.~Mora-Luna and R.~da Rocha,
  %``Dynamical tachyonic AdS/QCD and information entropy,''
  Phys.\ Lett.\ B {\bf 782} (2018) 607 
 % doi:10.1016/j.physletb.2018.06.007
  [{arXiv:1805.04485 [hep-th]}].

\bibitem{Braga:2018fyc} N.~R.~F.~Braga, L.~F.~Ferreira and R.~da Rocha,
  %\emph{Thermal dissociation of heavy mesons and configurational entropy,} 
  Phys.\ Lett.\ B {\bf 787} (2018) 16 
  [{arXiv:1808.10499 [hep-ph]}].

\bibitem{daSilva:2017jay} A.~Goncalves and R.~da Rocha,
  %``Information-entropic analysis of Korteweg?de Vries solitons in the quark?gluon plasma,''
  Phys.\ Lett.\ B {\bf 774} (2017) 98  
  %doi:10.1016/j.physletb.2017.09.046
  [{arXiv:1706.01482 [hep-ph]}].


\bibitem{Karapetyan:2018oye} G.~Karapetyan,
  %``Configurational entropy and $\rho$ and $\upphi$ mesons production in QCD,''
  Phys.\ Lett.\ B {\bf 781} (2018) 201 [{arXiv:1802.09105 [nucl-th]}].


\bibitem{Karapetyan:2018yhm}
  G.~Karapetyan,
  %``The nuclear configurational entropy approach to dynamical QCD effects,''
  Phys.\ Lett.\ B {\bf 787} (2018) 418 [arXiv:1807.04540 [nucl-th]].

\bibitem{Karapetyan:2016fai} G.~Karapetyan,
  %``Fine-tuning the Color-Glass Condensate with the nuclear configurational entropy,''
  EPL  {\bf 117} (2017) 18001 [{arXiv:1612.09564 [hep-ph]}].

\bibitem{Karapetyan:2017edu} G.~Karapetyan,
  %``The nuclear configurational entropy impact parameter dependence in the Color-Glass Condensate,''
  EPL {\bf 118} (2017) 38001 
  %doi:10.1209/0295-5075/118/38001
  [{arXiv:1705.10617 [hep-ph]}].

\bibitem{Ma:2018wtw} C.~W.~Ma, Y.~G.~Ma,
  %``Shannon Information Entropy in Heavy-ion Collisions,''
  Prog.\ Part.\ Nucl.\ Phys.\  {\bf 99} (2018) 120 [{arXiv:1801.02192 [nucl-th]}].

\bibitem{Gleiser:2013mga} M.~Gleiser and D.~Sowinski,
  %``Information-Entropic Stability Bound for Compact Objects: Application to Q-Balls and the Chandrasekhar Limit of Polytropes,''
  Phys.\ Lett.\ B {\bf 727} (2013) 272  [{arXiv:1307.0530 [hep-th]}].

\bibitem{Gleiser:2015rwa} M. Gleiser and N. Jiang, Phys.\ Rev.\ D {\bf 92} (2015) 044046  [{arXiv:1506.05722 [gr-qc]}].

\bibitem{Casadio:2016aum} R.~Casadio and R.~da Rocha,
  %``Stability of the graviton Bose-Einstein condensate in the brane-world,''
   Phys. Lett. B {\bf 763} (2016) 434 [{arXiv:1610.01572 [hep-th]}].

\bibitem{roldao} R.~A.~C.~Correa and R.~da Rocha,
 %``Configurational entropy in brane-world models,''
 Eur.\ Phys.\ J.\ C {\bf 75} (2015) 522.

\bibitem{Correa:2016pgr} R.~A.~C.~Correa, D.~M.~Dantas, C.~A.~S.~Almeida and R.~da Rocha,
  %``Bounds on topological Abelian string-vortex and string-cigar from information-entropic measure,''
  Phys.\ Lett.\ B {\bf 755} (2016) 358  
  [{arXiv:1601.00076 [hep-th]}].

\bibitem{Alves:2017ljt} A.~Alves, A.~G.~Dias,  R.~da Silva,
  %``Maximum Entropy Inferences on the Axion Mass in Models with Axion-Neutrino Interaction,''
 Braz. J. Phys. {\bf 47} (2017) 426 [{arXiv:1703.02061 [hep-ph]}].

\bibitem{Alves:2014ksa} A.~Alves, A.~G.~Dias, R.~da Silva,
  %``Maximum Entropy Principle and the Higgs Boson Mass,''
  Physica {\bf 420} (2015) 1 [{arXiv:1408.0827 [hep-ph]}].

\bibitem{Braga:2015jca} N.~R.~F.~Braga, M.~A. M.  Contreras and S.~Diles,
  %``Decay constants in soft wall AdS/QCD revisited,''
  Phys.\ Lett.\ B {\bf 763} (2016) 203 
  %doi:10.1016/j.physletb.2016.10.046
  [{arXiv:1507.04708 [hep-th]}].

\bibitem{Colangelo:2008us} P.~Colangelo, F.~De Fazio, F.~Giannuzzi, F.~Jugeau and S.~Nicotri,
 % ``Light scalar mesons in the soft wall model of AdS/QCD,''  
 Phys.\ Rev.\ D {\bf 78} (2008) 055009 [{arXiv:0807.1054 [hep-ph]}].  %%CITATION = ARXIV:0807.1054;%%  %86 citations counted in INSPIRE as of 14 Mar 2013

\bibitem{Afonin:2012jn} S.~S.~Afonin,
Phys. Lett. B {\bf 719} (2013) 399 [{arXiv:1210.5210 [hep-ph]}].

\bibitem{Rougemont:2017tlu}
  R.~Rougemont, R.~Critelli, J.~Noronha-Hostler, J.~Noronha and C.~Ratti,
  %``Dynamical versus equilibrium properties of the QCD phase transition: A holographic perspective,''
  Phys.\ Rev.\ D {\bf 96} (2017)  014032
 % doi:10.1103/PhysRevD.96.014032
  [arXiv:1704.05558 [hep-ph]].

\bibitem{dePaula} W.~de Paula, T.~Frederico, H.~Forkel and M.~Beyer,
  %``Dynamical AdS/QCD with area-law confinement and linear Regge trajectories,''
  Phys.\ Rev.\ D {\bf 79} (2009) 075019
  %doi:10.1103/PhysRevD.79.075019
  [{arXiv:0806.3830 [hep-ph]}].

\bibitem{dePaula:2009za} W.~de Paula and T.~Frederico,
  %``Scalar mesons within a dynamical holographic QCD model,''
  Phys.\ Lett.\ B {\bf 693} (2010) 287
 % doi:10.1016/j.physletb.2010.08.045
  [{arXiv:0908.4282 [hep-ph]}].

\bibitem{Capossoli:2016ydo} E.~Folco Capossoli, D.~Li and H.~Boschi-Filho,
  %``Dynamical corrections to the anomalous holographic soft wall model: the pomeron and the odderon,''
  Eur.\ Phys.\ J.\ C {\bf 76} (2016) 320 
  %doi:10.1140/epjc/s10052-016-4171-0
  [{arXiv:1604.01647 [hep-ph]}].

\bibitem{Ihl:2010zg} M.~Ihl, M.~Torres, H.~Boschi-Filho and C.~A.~B.~Bayona,
  %``Scalar and vector mesons of flavor chiral symmetry breaking in the Klebanov-Strassler background,''
  JHEP {\bf 1109} (2011) 026 [{arXiv:1010.0993 [hep-th]}].

\bibitem{Batell:2008me} B.~Batell, T.~Gherghetta and D.~Sword,
  %``The soft wall Standard Model,''
  Phys.\ Rev.\ D {\bf 78} (2008) 116011
  %doi:10.1103/PhysRevD.78.116011
  [{arXiv:0808.3977 [hep-ph]}].

\bibitem{BallonBayona:2007qr} C.~A.~Ballon Bayona, H.~Boschi-Filho and N.~R.~F.~Braga,
  %``Deep inelastic scattering from gauge string duality in the soft wall model,''
  JHEP {\bf 0803} (2008) 064
  %doi:10.1088/1126-6708/2008/03/064
  [{arXiv:0711.0221 [hep-th]}].



\bibitem{xu} F. Xu and M. Huang, 
%?Electric and magnetic screenings of gluons in a model with dimension-2 gluon condensate,? 
Chin. Phys. C {\bf 37} (2013) 014103 [arXiv:1111.5152 [hep-ph]].

\bibitem{Cherman:2008eh} A.~Cherman, T.~D.~Cohen and E.~S.~Werbos,
  %``The Chiral condensate in holographic models of QCD,''  
  Phys.\ Rev.\ C {\bf 79} (2009) 045203  [{arXiv:0804.1096 [hep-ph]}].  %%CITATION = ARXIV:0804.1096;%%  %32 citations counted in INSPIRE as of 08 Apr 2013

\bibitem{EKSS2005} J.~Erlich, E.~Katz, D.~T.~Son and M.~A.~Stephanov,
%``QCD and a holographic model of hadrons,''
Phys.\ Rev.\ Lett.\ \textbf{95} (2005) 261602 [{arXiv:hep-ph/0501128}].

\bibitem{Huang:2007fv} S.~He, M.~Huang, Q.~S.~Yan and Y.~Yang,
  %``Confront Holographic QCD with Regge Trajectories,''
  Eur.\ Phys.\ J.\ C {\bf 66} (2010) 187
  %doi:10.1140/epjc/s10052-010-1239-0
  [arXiv:0710.0988 [hep-ph]].

\bibitem{Braga:2016oem} N.~R.~F.~Braga and L.~F.~Ferreira,
  %``Thermal width of heavy quarkonia from an AdS/QCD model,''
  Phys.\ Rev.\ D {\bf 94} (2016)  094019
  % doi:10.1103/PhysRevD.94.094019
  [{arXiv:1606.09535 [hep-th]}].


\end{thebibliography}
\end{document}